\documentclass[a4paper,11pt]{article}
\pdfoutput=1

\usepackage{jheppub}

\usepackage[utf8]{inputenc}
\usepackage[T1]{fontenc}
\usepackage[english]{babel}
\usepackage{layout}
\usepackage{setspace}
\usepackage{lmodern}
\usepackage{enumitem}
\usepackage{soul}
\usepackage{eurosym}
\usepackage{url}

\usepackage{verbatim}
\usepackage{moreverb}
\usepackage{listings}
\usepackage{fancyhdr}

\usepackage{wrapfig}
\usepackage{graphicx}
\usepackage{pdfpages}

\usepackage{sectsty}

\usepackage{amsthm}
\usepackage{mathrsfs}
\usepackage{cancel}

\usepackage{makeidx}
\usepackage{titlesec}
\usepackage{multirow}
\usepackage{stmaryrd}

\usepackage{array}
\usepackage{booktabs}
\usepackage{multirow}
\usepackage{diagbox}

\usepackage{amsmath}

\usepackage{floatrow}
\newfloatcommand{capbtabbox}{table}[][\FBwidth]

\usepackage{enumitem}

\usepackage{footnote}

\newtheorem{Definition-Proposition}{Definition-Proposition}

\DeclareFontFamily{U}{mathx}{\hyphenchar\font45}
\DeclareFontShape{U}{mathx}{m}{n}{
      <5> <6> <7> <8> <9> <10>
      <10.95> <12> <14.4> <17.28> <20.74> <24.88>
      mathx10
      }{}
\DeclareSymbolFont{mathx}{U}{mathx}{m}{n}
\DeclareFontSubstitution{U}{mathx}{m}{n}
\DeclareMathAccent{\widebar}{0}{mathx}{"73}

\newcommand\varpm{\mathbin{\vcenter{\hbox{%
  \oalign{\hfil$\scriptstyle+$\hfil\cr
          \noalign{\kern-.3ex}
          $\scriptscriptstyle({-})$\cr}%
}}}}
\newcommand\varmp{\mathbin{\vcenter{\hbox{%
  \oalign{$\scriptstyle({+})$\cr
          \noalign{\kern-.3ex}
          \hfil$\scriptscriptstyle-$\hfil\cr}%
}}}}

\title{Rigorous treatment of the $\mathcal{S}^1 / \mathbb{Z}_2$ orbifold model with brane-Higgs couplings}

\author{Ruifeng Leng,}
\author{Gr\'egory Moreau,}
\author{Florian Nortier}

\affiliation{Université Paris-Saclay, CNRS/IN2P3, IJCLab, 91405 Orsay, France}
\emailAdd{ruifeng.leng@ijclab.in2p3.fr}
\emailAdd{moreau@ijclab.in2p3.fr}
\emailAdd{nortier@ijclab.in2p3.fr}

\abstract{We build rigorously the attractive five-dimensional model 
where bulk fermions propagate along the $ \mathcal{S}^1 / \mathbb{Z}_2$ orbifold and interact with a Higgs boson
localised at a fixed point of the extra dimension. The analytical calculation of the fermion mass spectrum and effective Yukawa couplings is shown to require
the introduction of either Essential Boundary Conditions (EBC) imposed by the model definition or certain Bilinear Brane Terms (BBT) in the action, instead of the  
usual brane-Higgs regularisations. The obtained fermion profiles along the extra dimension turn out to undergo some discontinuities, in particular at the
Higgs brane, which can be mathematically consistent if the action is well written with improper integrals. We also show that the $\mathbb{Z}_2$ parity 
transformations in the bulk do not affect the fermion chiralities, masses and couplings, in contrast with the EBC and the BBT, but when extended to the fixed points, 
they can generate the chiral nature of the theory and even select the Standard Model chirality set-up while fixing as well the fermion masses and couplings. 
Thanks to the strict analysis developed, the duality with the interval model is scrutinised.}

\begin{document} 
\maketitle
\flushbottom

\section{Introduction}
\label{Introduction}

As it is well known since the 2000s, the paradigm of models with additional spatial dimensions~\footnote{Together with the composite Higgs models which are dual models 
via the AdS/CFT correspondance.} constitutes an attractive alternative to supersymmetry for addressing the Standard Model (SM) puzzle of the gauge {\it hierarchy}. 
Furthermore, the warped dimension framework~\cite{Randall:1999ee} with SM fermions in the whole bulk~\cite{Chang:1999nh} offers an elegant geometrical principle of fermion 
profile overlap generating the SM fermion mass {\it hierarchy}~\cite{Gherghetta:2000qt} (see concrete application models {\it e.g.} in  
Ref.~\cite{Huber:2001ug,Casagrande:2008hr,Chang:2005ya,Moreau:2005kz,Moreau:2006np,Bouchart:2008vp}). In order 
to realise these two hierarchical features, the Brout-Englert-Higgs scalar field~\cite{Englert:1964et,Higgs:1964ia}, which is at the origin of the SM particle 
masses through the electroweak symmetry breaking, must be stuck at the so-called TeV-brane~\footnote{Let us 
mention here other possible phenomenological motivations, as from neutrino mass models, for the Higgs boson to be stuck at the boundary of an 
interval~\cite{Dienes:1998sb,Abada:2006yd,Grossman:1999ra,Huber:2002gp,Moreau:2004qe} or for fermions to propagate in the bulk~\cite{Frere:2003hn,Nortier:2020lbs}.} (or 
located in the bulk with a wave function strongly peaked at this brane). The TeV-brane is a 3-brane (three spatial dimensions) possibly at a boundary of the finite warped
extra dimension~\footnote{See for instance 
Ref.~\cite{Ledroit:2007ik,Bouchart:2009vq,Djouadi:2006rk,Djouadi:2007eg,Djouadi:2007fm,Djouadi:2009nb,Djouadi:2011aj,Angelescu:2017jyj} for its phenomenology 
and Ref.~\cite{Bouchart:2011va} in a supersymmetric context.}. 
More generally, a brane is an hypersurface located in an higher-dimensional space. It can arise in the context of string theories as D-branes which 
are dynamical objects with quantum properties~\cite{Polchinski:1996na,Bachas:1998rg} (see also Ref.~\cite{Aharony:1999ti,Duff:1996zn} for the supergravity limit of 
string theories)~\footnote{See Ref.~\cite{Fichet:2019owx,Nortier:2020vge} for brane-world effective field theories.}.

In this paper, we will study the original version~\cite{Randall:1999ee} 
of the warped dimension scenario based on the ${\cal S}^1/ \mathbb{Z}_2$ orbifold~\cite{Dixon:1985jw,Dixon:1986jc} 
where the extra space is compactified on a circle respecting a spatial parity
of the Lagrangian~\footnote{An orbifold ${\cal O}$ being defined as an extra compact manifold ${\cal C}$ with so-called fixed points 
where the introduced spatial transformation (element from a discrete group $G$) -- letting the Lagrangian invariant -- is just equivalent to the identity. 
It is noted ${\cal O} = {\cal C}/G$ and possesses thus singularities, not like a smooth manifold~\cite{Quiros:2003gg,Choi:2020dws}.}. 
Focusing our attention on the subtle bulk fermion interactions with the brane-Higgs field localised at a fixed point, we will analyse the toy model with a
flat extra dimension and the minimal field content: the results obtained on the fermion-Higgs coupling structure are directly applicable to the realistic warped model.

We will clarify the treatment of the bulk fermion couplings to the brane-localised Higgs boson, within the ${\cal S}^1/ \mathbb{Z}_2$ orbifold background, 
by building rigorously the four-dimensional (4D)~\footnote{Including time.} effective Lagrangian of the minimal model, 
that is by calculating consistently the Kaluza-Klein (KK) tower spectrum of fermion mass eigenvalues and the 4D effective Yukawa couplings
(via the fermion wave functions along the extra dimension).

In particular, we will demonstrate that no brane-Higgs regularisation [like smoothing the Higgs Dirac peak] 
should be applied (not necessary and no theoretical argument for it) in contrast with the usual regularisation procedure of literature (see Ref.~\cite{Angelescu:2019viv} and references therein)
and that, instead, one must introduce either Essential Boundary Conditions (EBC) on 5D fields~\footnote{Directly imposed by the model definition, in contrast with the Natural Boundary 
Conditions (NBC) deduced from the action minimisation condition.}, originating from the $\mathbb{Z}_2$ symmetry, or equivalently some Bilinear Brane Terms (BBT) in the fundamental 
5D Lagrangian. 
%((as shown from physical inconsistency, (role of) $Z_2$ conseq.: j=0 (then EBC), matching => BBT to announce in Section 2 / EBC same argues but introduced in Section 3.2))
The exact matching of the fermion mass spectra derived respectively through 4D and 5D methods will be used in order to confirm our analytical results.
All those statements (except the 4D approach) hold as well for the free case i.e. without Yukawa interactions.

This necessity of the presence of EBC or BBT (terms with the same form as in Ref.~\cite{Angelescu:2019viv,Nortier:2020xms}), in the 4D or 5D approach,
has been found as well~\cite{Angelescu:2019viv} in the finite interval scenario (the higher-dimensional framework of the other warped model version)
with identical brane-Higgs couplings to bulk fermions: this conclusion confirms that a specific treatment is required for 
point-like interactions between bulk fermions and brane-Higgs bosons in higher-dimensional spaces.

Besides, we will strictly describe and work out the entire known duality: identical physical quantities, namely the mass eigenvalues and 4D effective Yukawa couplings, 
are obtained in the different ${\cal S}^1/ \mathbb{Z}_2$ orbifold scenario with the Higgs boson localised at a fixed point 
and finite interval geometrical set-up with the Higgs field stuck at a boundary.

The EBC and BBT (forms including signs) choices, which should originate from an Ultra-Violet (UV) completion of the theory,  
turn out to induce the chiral nature of the low-energy effective theory as well as realising the specific SM fermion chiralities.
%((as in paper 1))
%((link to UV yet described-underlined))
%((BBT unicity over all possible choices => chirality prediction TH))
%((consider also close model where fields related at different boundaries à la Gr. or Lag vary under Z2
%  from (unknown) UV hypothesis to relate field variations))
Indeed, all these chirality properties are in fact not selected by the remaining sign choices for the 5D fields transformed via the spatial $\mathbb{Z}_2$ group -- as 
the solutions we find within this orbifold configuration can exhibit twist transformations (sign modification here) of the 5D fields, \`a la Scherk-Schwarz~\cite{Scherk:1979zr,Cremmer:1979uq}, 
through the extra space reflection.  
%((SS name for translation and non-orbifold:))  
We will even show that the transformation sign choices are just mathematical conventions without physical impacts on the SM field chiralities, 
the fermion mass spectrum and the 4D effective Yukawa couplings.

Nevertheless, in order to clarify the chirality aspects, we will also study a different scenario -- considered for example in Ref.~\cite{Gherghetta:2000kr} -- where the 
$\mathbb{Z}_2$ transformation definitions on the fields cover as well the fixed points themselves. It turns out that the associated transformation sign choices precisely at these
fixed points constitute here additional EBC, noted EBC', that have the capacity to select some of the previous EBC and hence to fix the chirality set-up. Once more, 
the r\^ole of these EBC' can be played instead by certain of the above BBT. Interestingly, such an inclusive $\mathbb{Z}_2$ symmetry definition can induce by itself the 
chiral nature of the theory as well as the SM chirality distribution over the various fields. This origin for the whole chirality configuration is not offered within the
simpler interval model for instance. In the presence of brane-localised Yukawa couplings, such an inclusive $\mathbb{Z}_2$ scenario can only be treated through the 4D method.
The fermion masses and couplings are also affected by this inclusive $\mathbb{Z}_2$ symmetry.

%((consistent jumps or correct treat lead to possibility of custodians (sol without zero modes) in this orbifold (no need at least for $ZxZ'$ in 1D) free / and Yuk if time
%add interval free yet and Yuk if time (=>pheno) ))

The action integral definition and integral domain end-points will be treated carefully. 
In particular the decomposition of the action to introduce 
improper integrals will appear to be required in the presence of orbifold fixed points or point-like fermion-boson interactions
(not located at the boundary of a finite extra space like an interval). 
Within this new and appropriate approach of the specific points along the extra dimension of the orbifold, we find for the free or Yukawa case that 
%((render possible))
some of the obtained consistent solutions exhibit certain field jumps at these fixed points and localised-interaction point. 
This interesting result of the possible existence of consistent profile jumps stands against one's first intuition~\cite{Csaki:2003sh,Azatov:2009na}, but those jumps are only induced
by sign flipping and not by point-like changes of the absolute value of the wave function amplitudes. 
%((intuitively [from QM]: abs proba smooth and sign arbitrary))
%((and phys sign diff to interpret in QFT
% anyway with "impurity" seems ok and undiscrete space so infinitely small point ok
% + smooth phys just intuition in class/QM
% + math/phys rigorously ok))

The analysis of the present orbifold background with brane-localised fermion-scalar interactions, as well as the previous results~\cite{Angelescu:2019viv} 
on the interval background, show that generally speaking 
the action expression does not systematically contain all the information allowing to fully define the model: 
in particular some EBC may be used (in contrast, the BBT are terms in the action)
depending on the brane treatment adopted or on the UV completion of the theory (which could introduce the BBT).
%((table shows ingredient necessary and EBC alternative in 4D method))
%((pyramid shows EBC importance/general method and EBC beyond action as well))

\section{Minimal $\mathcal{S}^1/\mathbb{Z}_2$ consistent model}
\label{Mini_Model}

\subsection{Geometry and symmetries: the proper action}
\label{5D_Geo}

\begin{figure}[h]
\centering
\includegraphics[width=5cm]{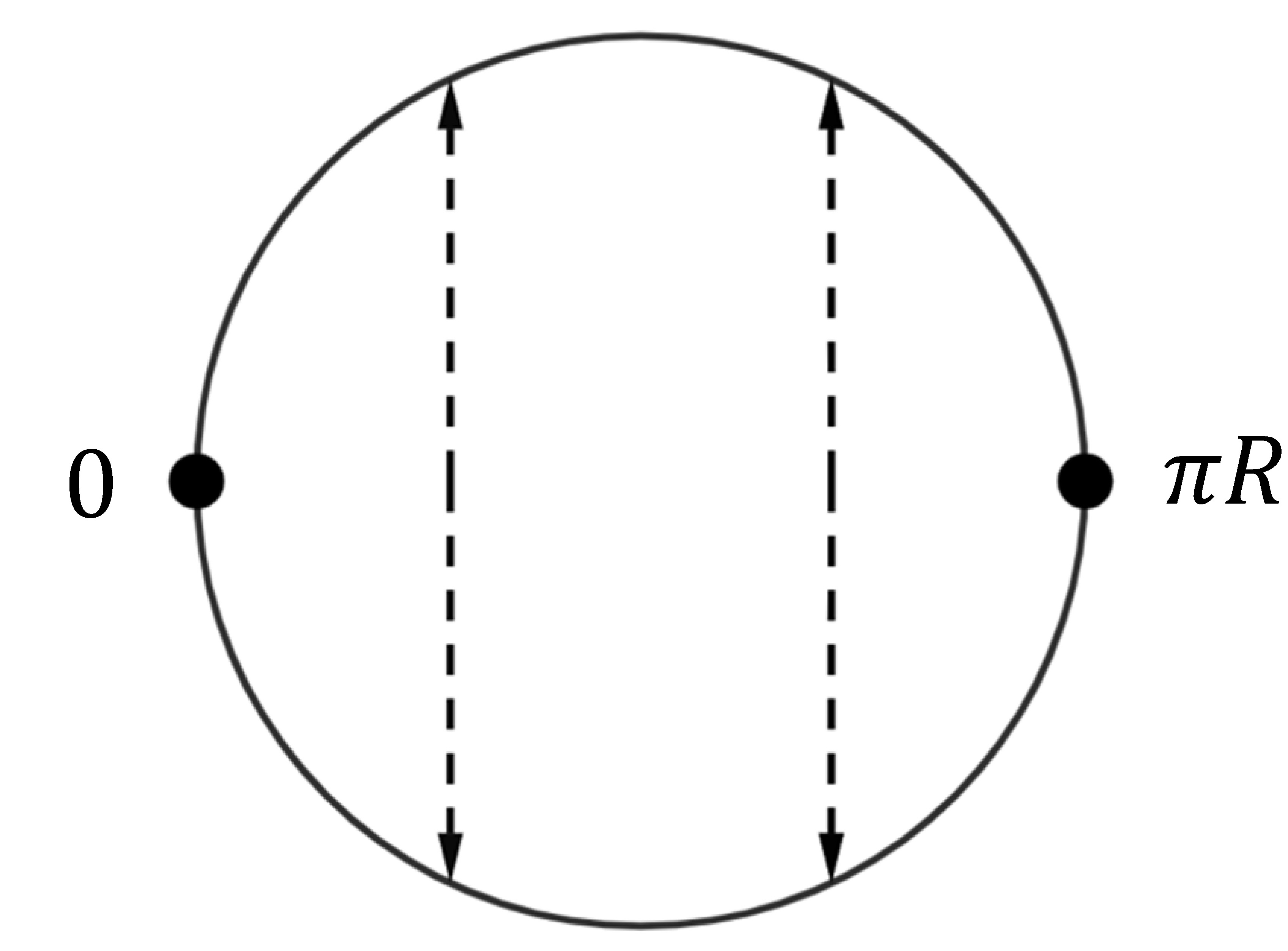}
\vspace{0.25cm}
\caption{$\mathcal{S}^1/ \mathbb{Z}_2$ orbifold picture. The fixed points at $y=0$ and $y=\pi R$ are indicated by the two black points.
The two examples of pairs of points with opposite coordinates, respectively indicated by the double dashed arrows, 
correspond to an identical Lagrangian density (for each pair).}
\label{S1_Z2_R}
\end{figure}

We consider the 5D space-time model with the product geometry $\mathcal{M}^4 \times \mathcal{S}^1/\mathbb{Z}_2$ described just below.
\begin{itemize}
\item $\mathcal{M}^4$ represents the usual 4D Minkowski space-time whose coordinates are denoted by $x^\mu$ where $\mu \in \llbracket 0, 3 \rrbracket$
is the Lorentz index of the covariant formalism. The metric conventions are given in Appendix~\ref{notations_and_conventions}.  
\item $\mathcal{S}^1/\mathbb{Z}_2$ stands for the extra space orbifold obtained from modding out the circle $\mathcal{S}^1$ by the discrete 
group~\footnote{Factor element, $e^{\pm i \frac{2 \pi}{2}}=-1$, and neutral element, $1$.} symmetry $\mathbb{Z}_2$.
\\ 
This circle $\mathcal{S}^1$ is characterised by a radius $R$ and its coordinate is $y \in (-\pi R, \pi R]$, not double-counting the point $y=\pi R$
since it is this point, by pure convention, that is chosen to be the junction point geometrically identified with the point $y=-\pi R$ (which we note: 
$-\pi R \equiv \pi R$)~\footnote{Another possible mathematical convention would have been, for instance, $-\frac{ 3\pi R}{2} \equiv \frac{\pi R}{2} $.}
in order to implement the circle periodicity. The circle could be constructed from the real axis by imposing a periodicity, that is by identifying geometrically 
an infinite number of translated regions of size $2 \pi R$ and hence by limiting the 1D space to the fundamental domain $(-\pi R,\pi R]$. 
\\
The (non-neutral) $\mathbb{Z}_2$ transformation on space, $y\to -y$~\footnote{The convention above, of having taken the coordinate origin at the strict
middle of the circle domain (or fundamental domain), renders the $\mathbb{Z}_2$ parity with respect to the origin more explicit and convenient to study.}, 
has a representation on a generic 5D field, 
\begin{equation}
\Phi(x^\mu, -y)={\mathcal T} \Phi(x^\mu, y) \, , \ \ \forall y \in (-\pi R, 0) \cup (0, \pi R) \, ,
\label{eq:5DFtrans}
\end{equation}
which must let the Lagrangian density invariant, by definition
of the symmetry: 
\begin{equation}
\mathcal{L}\left[\Phi(x^\mu, -y)\right] = \mathcal{L}\left[\Phi(x^\mu, y)\right] \, , \ \ \forall \, y \in (-\pi R, \pi R] \, .
\label{eq:Z2Lag}
\end{equation}
We mention that this equation can define a class of equivalence of a given coordinate $y_0$, defined as
$[y_0] = \{ y \in \mathcal{S}^1 \, \vert \, y \sim y_0 \}$ with $y \sim \pm y$, as illustrated symbolically on Fig.~\ref{S1_Z2_R}. 
\\
Two fixed points arise: $(y=0)\to (- 0=0)$ and $(y=\pi R)\to (- \pi R \equiv \pi R)$.
At these fixed points, the Lagrangian condition of Eq.~\eqref{eq:Z2Lag} is automatically satisfied:
$\mathcal{L}\left[\Phi(x^\mu, -0)\right] = \mathcal{L}\left[\Phi(x^\mu, 0)\right]$, and, 
$\mathcal{L}\left[\Phi(x^\mu, -\pi R)\right] = \mathcal{L}\left[\Phi(x^\mu, \pi R)\right]$,
so that ${\mathcal T}$ is naturally taken as the identity operator in Eq.~\eqref{eq:5DFtrans} since no transformation needs to apply on the fields there.
Another scenario will be analysed in Section~\ref{subsec:Z2inclusive}.
\end{itemize}

\vspace{0.35cm}

In order to properly write down the initial action, we urge the importance of taking care of possible field jumps along the extra dimension upon the reader.
We are going to show that the existence of a field jump in field theory can make sense mathematically if the action integration domain is properly divided 
at the jump location. Different discontinuity configurations must be considered. 
First, the hypothesis of a possible jump at any point of the bulk would lead to an infinite number of cuts in the action integration region which would obviously
not be treatable leading to unpredictable observables: this assumption is thus excluded.
Secondly, assuming an arbitrary finite number of possible jumps and hence of mathematical separations in the action domain, outside the fixed points, is not expected to 
affect the unique physical results -- like the fermion mass spectrum -- since none of those jump points exhibit some specific property: it is thus useless to explore this direction. 
Thirdly, the case of possible profile jumps at the two specific points that are the fixed points of the orbifold 
-- one of those two, $y=\pi R$, corresponding as well to the Yukawa coupling location 
(see Section~\ref{1_Yukawa}) -- remains to be studied. The effective presence of such profile jumps in some of the obtained solutions (see 
Figures~\ref{Profiles_S1/Z2_R_Free_Graphic} and \ref{Profiles_S1/Z2_R_Yuk_Graphic} respectively for the free and coupled fermion situations) confirms this possibility.  
For example, in case of a profile jump at $y=0$ (an identical discussion holds for the other fixed point at $y=\pi R$), regarding a well-defined Lagrangian 
integrand involving 5D fields over the whole action integration domain, we simply have to choose between the mathematical definitions of the left or right continuity 
for a generic profile function 
along the extra dimension: $f(0)=f(0^-)=\lim_{\epsilon\to 0} f(0-\epsilon)$ with $\epsilon > 0$, or, $f(0)=f(0^+)$. This choice is conventional and hence 
cannot affect numerical results, so let us choose conveniently 
\begin{align}
f(0)=f(0^+) \ \ \mbox{and} \ \ f(\pi R^-)=f(\pi R)
\label{eq:defFP} 
\end{align}
throughout this paper, in case of jumps at the fixed points. Then, the well-defined global 
action of this model must be written as a sum of some brane terms, 
an improper integral and a standard integration over different regions covering the whole physical domain of the circle:
\begin{align}
S_{\rm 5D} = S_{bulk} + S_{branes} \, , \ \mbox{with,} \ \ S_{bulk} \, = \, \int d^4x \left ( \int_{-\pi R^+}^{0^-} dy\ \mathcal{L}_{kin} + \int_0^{\pi R} dy\ \mathcal{L}_{kin} \right ) ,
\ \mbox{and,} \nonumber \\
\int d^4x \int_{-\pi R^+}^{0^-} dy\ \mathcal{L}_{kin} \  
\hat = \,  \lim_{a\to 0^-,\, b\to -\pi R^+} \, \int d^4x  \int_{b}^{a} dy\ \mathcal{L}_{kin}  \,
= \,  \lim_{\epsilon\to 0} \, \int d^4x  \int_{-\pi R+\epsilon}^{0-\epsilon} dy\ \mathcal{L}_{kin}  \, ,
\label{eq:Sstart} 
\end{align}
where $\epsilon > 0$,  
$S_{branes}$ represents action terms located at the orbifold fixed points and $\mathcal{L}_{kin}$ stands for the fermion kinetic terms of the Lagrangian density
(see next subsection).
Indeed, all the obtained fields will be well-defined at the two fixed points via Eq.~\eqref{eq:defFP}. Besides, for $\mathcal{L}_{kin}$ to be integrable over the entire region 
$y \in [0, \pi R]$, this Lagrangian density, which will involve profile derivatives $f'(y)$, must be well-defined over this region. The necessary (but not sufficient) condition for 
this last feature is that the profiles $f(y)$ have to be continuous on $[0, \pi R]$ and Eq.~\eqref{eq:defFP} also guarantees this continuity. 
For the consistency of the other integration term in Eq.~\eqref{eq:Sstart}, the profile continuity along $y \in [-\pi R^+,0^-]$~\footnote{To be clear, the integration 
domain $[-\pi R^+,0^-]$ corresponds to the spatial region along the extra dimension $]-\pi R,0[ \ \Leftrightarrow$ 
$(-\pi R,0)$ -- respectively the Francophone and Anglophone notations -- which does not include the fixed points at $y=0$ and $y=-\pi R$.} obviously reads as:
$$
\lim_{\kappa\to 0} f(0^- -\kappa)=\lim_{\epsilon\to 0} \lim_{\kappa\to 0} f([0-\epsilon] -\kappa)=\lim_{\omega \to 0} f(0 -\omega) =f(0^-)  \ ,
$$
with $\kappa > 0$, $\epsilon > 0$, $\omega = \epsilon + \kappa$ and similar equalities hold at the other region boundary $y=-\pi R^+$.
Furthermore, the worked out solutions $f(y)$ will (well) be derivable over the two regions $[-\pi R^+,0^-]$ and $[0, \pi R]$ 
(see Sections~\ref{S1/Z2_R_Free_EBC__Func} and \ref{S1/Z2_R_Yuk_BBT__Func} respectively for the free and coupled fermion situations) 
so that $\mathcal{L}_{kin}$ will be well-defined. For example, $f(y)$ is derivable in the region $[0, \pi R]$ at $y=0$ if and only if $f(y)$ is right-derivable at $y=0$, 
and the corresponding right-derivative does not diverge thanks to the first equality of Eq.~\eqref{eq:defFP}.
Notice that from the point of view of the integration by pieces of the action in Eq.~\eqref{eq:Sstart} precisely over the physical domain,  
the inclusion (or not) of the single points at $y=0$ or $y=\pi R\equiv - \pi R$ does not affect the integral results -- 
given the continuous form of the even $\mathcal{L}_{kin}$ over the two regions -- so that only consistent action definition arguments were considered here.

Finally, the Lagrangians of the whole expression~\eqref{eq:Sstart} 
will respect the $\mathbb{Z}_2$ symmetry since the Lagrangian $\mathcal{L}_{kin}$ will fulfill the condition~\eqref{eq:Z2Lag}
and the brane action will exclusively involve Lagrangians taken at fixed points like for example [see Eq.~\eqref{S_Yuk_S1/Z2_R__Func}], 
$S_{branes} \ni S_Y = \int d^4x \, \mathcal{L}_Y (x^\mu,\pi R)$.

\subsection{Field content and complete action}
\label{Model_Ex}

\subsubsection{Bulk fermion fields}
\label{1_Bulk Fermions}

Let us introduce the minimal spin-1/2 field content which allows to write down a SM Yukawa-like coupling between zero mode
fermions (of different chiralities) and a spin-0 field (see Section~\ref{1_Yukawa}). It is constituted by  
a pair of fermion fields called $Q$ and $D$. Those particles propagate along the circle $\mathcal{S}^1$, as we have in mind an 
extension of this toy model to a realistic scenario with bulk matter
({\it c.f.} Section~\ref{1_brTermSec}) where $Q,D$ will represent respectively the ${\rm SU(2)_L}$ 
gauge doublet down-component quark and the singlet down-quark.

The 5D fields $Q(x^\mu, y)$ and $D(x^\mu, y)$ -- of mass dimension 2 -- have the following kinetic terms [entering Eq.~\eqref{eq:Sstart}] 
which allow to recover canonical covariant kinetic terms for the associated fermions in the 4D effective action (as imposed
by the argument of decoupling limit~\footnote{From the theoretical consistency and phenomenological points of view,
the SM must be approximately recovered at low-energies in the limit of infinitely heavy KK excitations.}):
\begin{align}
\mathcal{L}_{kin} =  \frac{i}{2} \ \left( \bar{Q}\Gamma^M \overleftrightarrow{\partial_M} Q + \bar{D}\Gamma^M \overleftrightarrow{\partial_M} D \right) \ ,
\label{S_Psi_S1/Z2_R__Func}
\end{align}
using the standard notations 
$\overleftrightarrow{\partial_M}=\overrightarrow{\partial_M}-\overleftarrow{\partial_M}$, $\partial_M=\partial/\partial x^M$,
$x^M =(x^\mu, y)$ with $M \in \llbracket 0, 4 \rrbracket$ for the coordinates $x^M \in \mathcal{M}^4 \times \mathcal{S}^1/\mathbb{Z}_2$
and $\Gamma^M$ for the 5D Dirac matrices ({\it c.f.} Appendix~\ref{notations_and_conventions}).
In the used conventions, the 5D Dirac spinor, being in the irreducible representation of the Lorentz group, reads for example for $Q$ as,
\begin{equation}
Q = \mathcal{Q}_L + \mathcal{Q}_R
\ \ \text{with} \ \ 
\mathcal{Q}_L =
\begin{pmatrix}
Q_L \\
0
\end{pmatrix}
\ \ \text{,} \ \ 
\mathcal{Q}_R =
\begin{pmatrix}
0 \\
Q_R
\end{pmatrix} \  ,
\label{1_5DDiracSp}
\end{equation}
in terms of the two two-component Weyl spinors $Q_L$, $Q_R$, $L/R$ standing for the Left/Right chirality,
and as usually $\bar{Q} = Q^\dagger \gamma^0$.

As stated at the end of previous section, the Lagrangian 
$\mathcal{L}_{kin}$ must obey the condition~\eqref{eq:Z2Lag}. For this purpose, the $\mathbb{Z}_2$ transformation~\eqref{eq:5DFtrans}
on the 5D fields $Q(x^\mu, y)$ and $D(x^\mu, y)$ can take four different forms which constitute Essential Conditions (EC) issued from the model definition:
\begin{equation} 
Type \,\uppercase\expandafter{\romannumeral1}
\left\{
\begin{array}{c c c}
Q \left( x^\mu, -y \right) = - \gamma^5 \, Q \left( x^\mu, y \right) \, \Longrightarrow \, Q_L \ \text{even}, \ Q_R \ \text{odd},
\\ \vspace{-0.2cm} \\
D \left( x^\mu, -y \right) = \gamma^5 \, D \left( x^\mu, y \right) \, \Longrightarrow \, D_L \ \text{odd}, \ D_R \ \text{even},
\end{array}
\right.
\label{Parity_S1/Z2_R_1}
\end{equation}
\begin{equation} 
Type \,\uppercase\expandafter{\romannumeral2}
\left\{
\begin{array}{c c c}
Q \left( x^\mu, -y \right) = \gamma^5 \, Q \left( x^\mu, y \right) \, \Longrightarrow \, Q_L \ \text{odd}, \ Q_R \ \text{even},
\\ \vspace{-0.2cm} \\
D \left( x^\mu, -y \right) = - \gamma^5 \, D \left( x^\mu, y \right) \, \Longrightarrow \, D_L \ \text{even}, \ D_R \ \text{odd},
\end{array}
\right.
\label{Parity_S1/Z2_R_2}
\end{equation}
\begin{equation} 
Type \,\uppercase\expandafter{\romannumeral3}
\left\{
\begin{array}{c c c}
Q \left( x^\mu, -y \right) = - \gamma^5 \, Q \left( x^\mu, y \right) \, \Longrightarrow \, Q_L \ \text{even}, \ Q_R \ \text{odd},
\\ \vspace{-0.2cm} \\
D \left( x^\mu, -y \right) = - \gamma^5 \, D \left( x^\mu, y \right) \, \Longrightarrow \, D_L \ \text{even}, \ D_R \ \text{odd},
\end{array}
\right.
\label{Parity_S1/Z2_R_3}
\end{equation}
\begin{equation} 
Type \,\uppercase\expandafter{\romannumeral4}
\left\{
\begin{array}{c c c}
Q \left( x^\mu, -y \right) = \gamma^5 \, Q \left( x^\mu, y \right) \, \Longrightarrow \, Q_L \ \text{odd}, \ Q_R \ \text{even},
\\ \vspace{-0.2cm} \\
D \left( x^\mu, -y \right) = \gamma^5 \, D \left( x^\mu, y \right) \, \Longrightarrow \, D_L \ \text{odd}, \ D_R \ \text{even},
\end{array}
\right.
\label{Parity_S1/Z2_R_4}
\end{equation}
under which the Lagrangian~\eqref{S_Psi_S1/Z2_R__Func} is indeed invariant, as appears by using the properties of the $\gamma_5$ Dirac matrix 
and the odd parity of the fifth partial derivative $\partial_4$. Notice that the $\mathbb{Z}_2$ parity (second order cyclic group) does not allow complex phase factors
in the transformations: $$F|_{y} = e^{i \theta_F} \gamma^5 F|_{-y} = e^{i \theta_F} \gamma^5 e^{i \theta_F} \gamma^5 F|_{y} = \left(e^{i \theta_F} \right)^2 F|_{y} \, .$$
Using the $\gamma_5$ definition of Appendix~\ref{notations_and_conventions} together with Eq.~\eqref{1_5DDiracSp}, we already deduce some
information on the possible 5D chiral field parities with respect to $y=0$, as indicated in Eq.~\eqref{Parity_S1/Z2_R_1}-\eqref{Parity_S1/Z2_R_4}.

Based on Eq.~\eqref{1_5DDiracSp}, we can rewrite the bulk Lagrangian of Eq.~\eqref{S_Psi_S1/Z2_R__Func} in forms which are
convenient to see at a glance the Lagrangian even parity, simply by using the occurence of fixed 5D field parities, different for the
Left/Right chiralities [{\it c.f.} Eq.~\eqref{Parity_S1/Z2_R_1}-\eqref{Parity_S1/Z2_R_4}], and the $\partial_4$ odd parity:
\begin{align}
\mathcal{L}_{kin} &= \dfrac{1}{2} \left( i Q^\dagger_R \sigma^\mu \overleftrightarrow{\partial_\mu} Q_R + i Q^\dagger_L \bar{\sigma}^\mu \overleftrightarrow{\partial_\mu} Q_L - Q^\dagger_R \overleftrightarrow{\partial_4} Q_L + Q^\dagger_L \overleftrightarrow{\partial_4} Q_R \right) + \{ Q \leftrightarrow D \} \, , \nonumber \\
&= \dfrac{1}{2} \left( i \widebar{\mathcal{Q}}_R \gamma^\mu \overleftrightarrow{\partial_\mu} \mathcal{Q}_R + i \widebar{\mathcal{Q}}_L \gamma^\mu \overleftrightarrow{\partial_\mu} \mathcal{Q}_L - \widebar{\mathcal{Q}}_R \overleftrightarrow{\partial_4} \mathcal{Q}_L + \widebar{\mathcal{Q}}_L \overleftrightarrow{\partial_4} \mathcal{Q}_R \right) + \{ \mathcal{Q} \leftrightarrow \mathcal{D} \} \, ,
\nonumber%\label{1_L_2}
\end{align}
where the low double arrows indicate a replacement of 5D fields in the previous terms and the matrices $\sigma^\mu, \bar{\sigma}^\mu$ are defined in
Appendix~\ref{notations_and_conventions}.

\subsubsection{Brane-localised scalar field}
\label{1_Brane Localized Scalar Field}

The questions about the mass calculation arise when the bulk fermions couple to a single 
4D real scalar field $H$ (mass dimension 1) which is confined at a fixed point of the orbifold, as in the studied model
(inspired by the warped scenario addressing the gauge hierarchy problem). We simply choose this fixed point to be at $y=\pi R$, 
rather than $y=0$, which is a purely mathematical convention since these two points belong to a circle. The real scalar field has an action of the generic form,
\begin{align}
&S_H  = \int d^4x \left [ \, \dfrac{1}{2} \, \partial_\mu H \partial^\mu H - V(H) \right ] \, ,   
\label{1_eq:actionH}
\end{align}
with a potential $V$ possessing a minimum which generates a non-vanishing Vacuum Expectation Value (VEV) for the field $H$ expanded as
\begin{equation}
H(x^\mu)=\frac{v+h(x^\mu)}{\sqrt{2}} \, ,
\label{1_H_exp}
\end{equation}
in analogy with the SM Higgs field.

\subsubsection{Yukawa interactions}
\label{1_Yukawa}

We consider the following Yukawa interactions allowing to study the subtleties induced by the coupling of the above brane-scalar field (at $y=\pi R$)
to the introduced bulk fermions,
\begin{align}
&S_{Y} = \int d^4x \,  \mathcal{L}_Y (x^\mu,\pi R) \, , \text{with,} \ \ 
\mathcal{L}_Y = - Y_5\ HQ^\dagger_LD_R - Y^\prime_5\ HQ^\dagger_RD_L +  {\rm H.c.} \, .
\label{S_Yuk_S1/Z2_R__Func}
\end{align}
Notice that considering operators involving the fields $H$, $Q$, $D$ up to dimension 5 allows to include such a Yukawa coupling. 
Let us recall here that in case of profile jumps at the fixed point at $y=\pi R$, the 5D fields $Q_{L/R}(x^\mu,\pi R)$, $D_{L/R}(x^\mu,\pi R)$ 
are defined through the profile convention~\eqref{eq:defFP}, as already described. 
The studied model with a Yukawa coupling at a fixed point will turn out to be dual to the
interval model including a Yukawa coupling at a boundary (see Section~\ref{S1/Z2_R_LM__Func}).

The complex $Y_5= \text{e}^{i \alpha_Y} |Y_5|$ and $Y_5'= \text{e}^{i \alpha_Y'} |Y_5'|$ Yukawa coupling constants, 
entering Eq.~\eqref{S_Yuk_S1/Z2_R__Func}, are independent 
and a well-defined 4D chirality holds for the fermion fields on the 3-brane strictly at $y = \pi R$~\cite{Angelescu:2019viv,Azatov:2009na}.
To avoid the introduction of a new energy scale, in the spirit of the warped model,
we can define the 5D Yukawa coupling constants as 
\begin{align}
Y_5=y_4 \times 2 \pi R \, , \ {\rm and,} \ Y^\prime_5=y_4' \times 2\pi R \, , 
\label{eq:RefDecouplY5}
\end{align}
where $y_4$, $y_4'$ are dimensionless coupling constants of ${\cal O}(1)$. Then, 
$y_4$ can be approximately identified with the SM Yukawa coupling constant within the decoupling limit,
as will be described in Eq.~\eqref{MS_DL_Interval & L_N__Func_Yuk_5D}-\eqref{MS_DL_S1/Z2_R__Func_Yuk_5D_BBT}.

When calculating the tower of excited fermion masses, we restrict our considerations to the VEV of $H$ 
and concentrate our attention on the following part of the action~\eqref{S_Yuk_S1/Z2_R__Func},
\begin{align}
&S_{X} = \int d^4x \, \mathcal{L}_X (x^\mu,\pi R) \, , \text{with,} \ \ 
\mathcal{L}_X = - X\  Q^\dagger_L  D_R - X^\prime\  Q^\dagger_R D_L +  {\rm H.c.} \, ,
\label{S_X_S1/Z2_R__Func}
\end{align}
with the compact notations $X = v Y_5/\sqrt{2}$ and $X' = v Y_5' /\sqrt{2}$. Based on Eq.~\eqref{1_H_exp},
the complete action reads as, $S_Y = S_X + S_{int}$, with the localised fermion-scalar interaction terms:
\begin{align}
&S_{int} = \int d^4x\, \mathcal{L}_{int} (x^\mu,\pi R) \, , \text{with,} \ \ 
\mathcal{L}_{int} = - \dfrac{Y_5}{\sqrt{2}} \ h Q^\dagger_L  D_R - \dfrac{Y_5'}{\sqrt{2}} \ h  Q^\dagger_R D_L +  {\rm H.c.} \, ,
\label{S_h_S1/Z2_R__Func}
\end{align}
that allow to work out the 4D effective Yukawa coupling constants.

\subsubsection{Bilinear brane terms}
\label{1_brTermSec}

Introducing all the covariant operators up to mass dimension 5 [like for the Yukawa couplings~\eqref{S_Yuk_S1/Z2_R__Func}] in this model, 
one should consider as well the dimension 4 operators given just below, that we call the BBT like in Ref.~\cite{Angelescu:2019viv}.
Furthermore, the presence of the BBT has several justifications: 
(i) they allow to avoid physical consistency problems both in the free case 
(see Sections~\ref{S1/Z2_R_Free_NBC__Func} and \ref{S1/Z2_R_Free_BBT__Func}) and with Yukawa couplings 
(Sections~\ref{S1/Z2_R_Yuk_NBC__Func} and \ref{S1/Z2_R_Yuk_BBT__Func});
(ii) they play the r\^ole of defining well the model at the two orbifold fixed points both in the free case 
(see Sections~\ref{S1/Z2_R_Free_EBC__Func} and \ref{S1/Z2_R_Free_BBT__Func}) and with Yukawa couplings (Section~\ref{S1/Z2_R_Yuk_BBT__Func});
(iii) they induce the expected matching of the analytical results on the spectrum derived through the 4D and 5D approaches
(see Sections~\ref{S1/Z2_R_Yuk_4D__Func} and \ref{S1/Z2_R_Yuk_BBT__Func}).

The following BBT lead to the SM chirality configuration,
\begin{align}
S_B = \int d^4x \ \left( \sigma^Q_{0} \, \left. \bar{Q}Q \right|_{0} + \sigma^Q_{\pi R} \, \left. \bar{Q}Q \right|_{\pi R} 
+ \sigma^D_{0} \, \left. \bar{D}D \right|_{0} + \sigma^D_{\pi R} \, \left. \bar{D}D \right|_{\pi R} \, \right) \, , 
\label{S_BBT_S1/Z2_R__Func}
\end{align}
where $\sigma^Q_{0(\pi R)} =\varpm$, $\sigma^D_{0(\pi R)} = \varmp$ and for example $\left. \bar{Q}Q \right|_{0} = \bar{Q}(x^\mu,0)Q(x^\mu,0)$.
Indeed, without Yukawa couplings, 
these terms will induce only a non-vanishing profile $q_L^0(y)$ [see line~2 of Eq.~\eqref{completeEBCsm0} and Table~\ref{tab:SMQ&D_Free_Z2}
in case of the zero-mode with mass $m_0=0$]
in the 5D field $Q_L(x^\mu,y)$ so that only the Left-handed 4D field $Q^0_L(x^\mu)$ will exist. This zero-mode $Q^0_L(x^\mu)$, without KK mass contribution, 
constitutes the lightest mode of the KK tower and also the SM state. Hence, we can well recover the SM configuration: a chiral field content and a Left-handed 
4D field potentially representing the ${\rm SU(2)_L}$ quark doublet in the direct extension to gauge symmetries (and three flavours). Given that, similarly, the 
BBT~\eqref{S_BBT_S1/Z2_R__Func} will exclusively lead to a Right-handed 4D field $D^0_R(x^\mu)$ [line~1 of Eq.~\eqref{completeEBCsm0}] 
potentially representing the SM down quark type (gauge 
singlet). When adding the Yukawa couplings~\eqref{S_Yuk_S1/Z2_R__Func}, this SM chirality set-up remains though it is no more explicit due to the 
$Q^n(x^\mu)$-$D^n(x^\mu)$ mixing, via vector-like KK state mixings, which induces some vector-like mass eigenstates $\psi^0_{L/R}(x^\mu)$ for the lightest modes 
of the tower (see Sections~\ref{S1/Z2_R_Yuk_4D__Func} and \ref{S1/Z2_R_Yuk_BBT__Func}). In the decoupling limit where heavy KK state mixings tend to vanish, 
the SM chirality configuration is recovered as expected.

For completeness, let us underline that in the free case, the opposite BBT signs, 
$\sigma^Q_{0(\pi R)} =\varmp$, $\sigma^D_{0(\pi R)} = \varpm$, would lead to a chiral set-up for the zero-modes
but different from the potential SM chirality configuration, namely: $Q_R^0(x^\mu),D_L^0(x^\mu)$.
Similarly, the BBT signs, $\sigma^Q_{0(\pi R)} =\varpm$, $\sigma^D_{0(\pi R)} = \varpm$, would lead to the set-up, $Q_L^0(x^\mu),D_L^0(x^\mu)$,
and, $\sigma^Q_{0(\pi R)} =\varmp$, $\sigma^D_{0(\pi R)} = \varmp$, to, $Q_R^0(x^\mu),D_R^0(x^\mu)$.

Finally, as will be described in the Sections~\ref{S1/Z2_R_Free_EBC__Func} and \ref{S1/Z2_R_Free_BBT__Func}, the possible signs, 
$\sigma^Q_{0(\pi R)} =\pm$ (same sign for $0$ and $\pi R$), would instead lead to the profile solutions~\eqref{completeEBCcusto0} 
with two non-vanishing profiles for the lightest modes (as $m_0\neq 0$) and hence to vector-like states: $Q_{L/R}^{0}(x^\mu)$.
The same statement holds for $\sigma^D_{0(\pi R)} =\pm$ and thus $D_{L/R}^{0}(x^\mu)$. Such massive vector-like states~\footnote{Extensive 
phenomenology at colliders has been developed about such vector-like particles~\cite{Azatov:2012rj,Bonne:2012im,Moreau:2012da,Gopalakrishna:2013hua,Angelescu:2015kga}.} 
can be used to build custodially protected warped models~\cite{Agashe:2003zs} and are then called custodians (see for instance
Ref.~\cite{Bouchart:2008vp}). Of course there exist 8 remaining cases combining the above 
Lagrangian sign configurations: $\sigma^Q_{0(\pi R)} =\varpm, \varmp$, $\sigma^D_{0(\pi R)} = \pm$, and, 
$\sigma^Q_{0(\pi R)} =\pm$, $\sigma^D_{0(\pi R)} = \varpm, \varmp$.

Therefore, it appears clearly that the BBT control the chiral configurations of the model. The UV completion of the theory can be at the origin 
of the BBT and hence of the chirality set-up: chiral nature of the theory and specific chiralities of the various fields.
 
To end up this section, we note that the complete toy model studied is characterised by the action,
\begin{eqnarray}
S_{5D} = S_{bulk} + S_{branes} = S_{bulk} + S_H  + S_X + S_{int} + S_B \ . \label{1_eq:actionTot}
\end{eqnarray}
The conclusions that will be derived in the present work can be directly extended to the realistic warped model with SM bulk matter 
addressing the fermion mass and gauge hierarchies, along the same lines as the flavour and gauge symmetry generalisations described in details 
in the Section~2.6 of Ref.~\cite{Angelescu:2019viv}.

\section{Free bulk fermions on the orbifold}
\label{S1/Z2_R_Free__Func}

In this section, we calculate the fermionic mass spectrum for the free case where $Y_5=Y_5'=0$ in the action piece 
$S_Y$ given by Eq.~\eqref{S_Yuk_S1/Z2_R__Func}.

\subsection{Applying the NBC}
\label{S1/Z2_R_Free_NBC__Func}

We start by considering the bulk action part, $$S_{bulk} \, ,$$ of Eq.~\eqref{eq:Sstart}, from the considered action, $S_{5D}$, 
of Eq.~\eqref{1_eq:actionTot}. We apply the least action principle to it which leads to two relations of the kind, $\delta_{\bar{F}} S_{bulk} = 0$, one for each of the unknown 
5D fields $F=Q,D$, and two corresponding ones, $\delta_{F} S_{bulk} = 0$, involving the complex conjugate fields~\footnote{The 
equations of motion and boundary conditions derived from the least action principle for the fields and their conjugates are trivially related through 
the Hermitian conjugation.}, 
since the elementary field variations $\delta Q_{\alpha}$, 
$\delta\bar{Q}_{\alpha}$, $\delta D_{\alpha}$ and $\delta\bar{D}_{\alpha}$ (see Appendix~\ref{app:Spin.1}) are generic and hence independent from each other. 
Using compact notations, like for example, 
$$\sum_{\alpha=1}^4 \delta \bar{F}_\alpha \dfrac{\partial \mathcal{L}_{kin}}{\partial \bar{F}_\alpha} \, \hat = \, 
\delta {\bar{F}} \dfrac{\partial \mathcal{L}_{kin}}{\partial \bar{F}} \, ,$$ we can write in particular~\footnote{We omit the global 4-divergence which vanishes in the action 
integration due to vanishing fields at the boundaries at infinities. Indeed, when minimising the action, the varied terms must vanish separately at infinite boundaries, 
since the non-vanishing field variations at boundaries are independent from each other and from the bulk ones (see also Ref.~\cite{Peskin:1995ev}). 
This is realised by the local physics statement which induces vanishing fields at infinities due to the wave function normalisation conditions (see also 
Ref.~\cite{Schwartz:2013pla}).},
\begin{align}
\delta_{\bar{F}} S_{bulk} 
&= \displaystyle{ \int d^4x \, \left(\int_{-\pi R^+}^{0^-} + \int_{0}^{ \pi R}\right) dy \ \left\{ \delta {\bar{F}} \dfrac{\partial \mathcal{L}_{kin}}{\partial \bar{F}} 
+ \delta \left( \partial_M \bar{F} \right) \dfrac{\partial \mathcal{L}_{kin}}{\partial \, \partial_M \bar{F}} \right\}} \nonumber \\
&= \displaystyle{ \int d^4x \, \left(\int_{-\pi R^+}^{0^-} + \int_{0}^{ \pi R}\right) dy \ \left\{ \delta {\bar{F}} \dfrac{\partial \mathcal{L}_{kin}}{\partial \bar{F}} 
+ \partial_M \left[  \delta \bar{F}  \dfrac{\partial \mathcal{L}_{kin}}{\partial \, \partial_M \bar{F}} \right] 
- \delta \bar{F} \, \partial_M \dfrac{\partial \mathcal{L}_{kin}}{\partial \, \partial_M \bar{F}} \right\} } \nonumber 
\end{align}
\begin{align}
&= \displaystyle{ \int d^4x \, \left(\int_{-\pi R^+}^{0^-} + \int_{0}^{ \pi R}\right) dy \ \left\{ \delta {\bar{F}} \left[ \dfrac{\partial \mathcal{L}_{kin}}{\partial \bar{F}} 
- \partial_M \dfrac{\partial \mathcal{L}_{kin}}{\partial \, \partial_M \bar{F}} \right] \right\} } 
\nonumber \\ &  \ \ \ \ \ \ \ \ \ \ \ \ \ \ \ \ \ \ \ \ \ \ \ \ \ \ \ \ \ \ \
+ \displaystyle{ \int d^4x \ \left( \left. \delta {\bar{F}} \dfrac{\partial \mathcal{L}_{kin}}{\partial \, \partial_4 \bar{F}} \right|_{-\pi R^+}^{0^-} 
+ \left. \delta {\bar{F}} \dfrac{\partial \mathcal{L}_{kin}}{\partial \, \partial_4 \bar{F}} \right|_{0}^{ \pi R}\right)} \ .
\label{S_Psi_S1/Z2_R__Func_HVP}
\end{align}
Based on the Lagrangian $\mathcal{L}_{kin}$ of Eq.~\eqref{S_Psi_S1/Z2_R__Func}, these two bulk terms take the same form
(the first one being calculated explicitly in 
Eq.~\eqref{app:CompCalc} to clarify the spinor component treatment) and the two remaining brane terms can be calculated as well:
\begin{align}
\delta_{\bar{F}} S_{bulk} &= \displaystyle{ \int d^4x \, \left(\int_{-\pi R^+}^{0^-} + \int_{0}^{ \pi R}\right) dy \ \left\{ \delta {\bar{F}} \left[ i \Gamma^{M} \partial_{M} F \right] \right\} } 
\nonumber \\ & \ \ \ \ \ \ \ \ \ \ \ \ \ \ \ \ \ \ \ \ \ \ \ \ \ \ \ \ \ \ \
+ \displaystyle{ \int d^4x \ \left(\left. \delta {\bar{F}} \left[ -\dfrac{\gamma^5}{2} F \right] \right|_{0^+}^{\pi R^-} 
+ \left. \delta {\bar{F}} \left[ -\dfrac{\gamma^5}{2} F \right] \right|_{0}^{ \pi R}\right)} \ ,
\label{S_Psi_S1/Z2_R__Func_HVP_P}
\end{align}
where we have further invoked the $\mathbb{Z}_2$ transformations~\eqref{Parity_S1/Z2_R_1}-\eqref{Parity_S1/Z2_R_4} for the generic 5D field,
Eq.~\eqref{Parity_Delta_Sp_bar_S1/Z2_R} for its variation and $\gamma^5$ properties:
\begin{align}
\left. \delta {\bar{F}} \left[ - \dfrac{\gamma^5}{2} F \right] \right|_{0^-, -\pi R^+} &= \left. \left(\mp \delta \bar{F}  \gamma^5\right) 
\left[ - \dfrac{\gamma^5}{2} \left(\pm   \gamma^5 F\right) \right] \right|_{0^+, \pi R^-} = - \left. \delta {\bar{F}} 
\left[ - \dfrac{\gamma^5}{2} F \right] \right|_{0^+, \pi R^-} \ . \nonumber
\end{align}
Then thanks to Eq.~\eqref{eq:defFP}~\footnote{Those continuity relations lead to $\bar F\vert_0=\bar F\vert_{0^+}$, i.e. $\bar F_\alpha\vert_0=\bar F_\alpha\vert_{0^+}$
[{\it c.f.} Eq.~\eqref{Sp_bar_C2C}],   
and in turn to $\delta \bar F_\alpha\vert_0=\delta \bar F_\alpha\vert_{0^+}$ which can be written as $\delta \bar F\vert_0=\delta \bar F\vert_{0^+}$ via 
Eq.~\eqref{Sp_group}. Similarly we get $\delta \bar F\vert_{\pi R}=\delta \bar F\vert_{\pi R^-}$.} and Eq.~\eqref{Sp_group-2comp}-\eqref{Sp_group-2comp-bis}, respectively, 
the expression~\eqref{S_Psi_S1/Z2_R__Func_HVP_P} simplifies to,
\begin{align}
\delta_{\bar{F}} S_{bulk} 
&= \displaystyle{ \int d^4x \, \left(\int_{-\pi R^+}^{0^-} + \int_{0}^{ \pi R}\right) dy \ \left\{ \delta {\bar{F}} \left[i \Gamma^{M} \partial_{M} F \right] \right\} } 
+ \displaystyle{ \int d^4x \ 2\left. \delta {\bar{F}} \left[ -\dfrac{\gamma^5}{2} F \right] \right|_{0}^{ \pi R}} 
\nonumber \\ 
= & \displaystyle{ \int d^4x \, \left(\int_{-\pi R^+}^{0^-} + \int_{0}^{ \pi R}\right) dy \ \left\{ \delta {\bar{F}} \left[i \Gamma^{M} \partial_{M} F \right] \right\} } 
+ \displaystyle{ \int d^4x 
\left. \left[ \delta {F^\dagger_R} F_L - \delta {F^\dagger_L} F_R \right] \right|_{0}^{ \pi R} } .  
\label{S_Psi_BT_S1/Z2_R__Func_HVP_P}
\end{align}
In this expression, the bulk and brane variations -- respectively the volume and surface terms -- must vanish separately due to independent
field variations (no reason to be linked). Besides all those field variations are not vanishing (unknown fields) so that we get the bulk Equations Of Motion (EOM),
\begin{equation} 
i \, \Gamma^{M} \partial_{M} F=0 \, ,
 \ \ \ \ \forall \, x^\mu, \ \forall \, y \in [-\pi R^+, 0^-] \cup [0, \pi R] \ ,
\label{ELE_S1/Z2_R__Func_Free_Chiral}
\end{equation}
and the Natural Boundary Conditions (NBC),
\begin{equation}
F_L|_{0} = F_R|_{0} = F_L|_{\pi R} = F_R|_{\pi R} = 0 \, .
\label{BC_S1/Z2_R__Func_Free_NBC}
\end{equation}
At this level, we can first solve Eq.~\eqref{ELE_S1/Z2_R__Func_Free_Chiral} together with Eq.~\eqref{BC_S1/Z2_R__Func_Free_NBC} to find out the $F$ fields 
over the domain, $y \in [0, \pi R]$. This is precisely what has been done in the preliminary Ref.~\cite{Angelescu:2019viv} where the two exactly identical Eq.~(3.3) 
and (3.4) [there] have been solved over the interval, $y \in [0, L]$. Since the fields are continuous over $y \in [0, \pi R]$ [{\it c.f.} Eq.~\eqref{eq:defFP}]
like there over $y \in [0, L]$, we can thus apply here the results obtained in this reference: the solutions found for  
Eq.~\eqref{ELE_S1/Z2_R__Func_Free_Chiral}-\eqref{BC_S1/Z2_R__Func_Free_NBC} are expressed through the KK decomposition (with a similar choice
of global factor),
\begin{equation}
F_{L/R} \left( x^\mu, y \right) = \dfrac{1}{\sqrt{2 \pi R}} \displaystyle{ \sum_{n=0}^{+\infty} f^n_{L/R}(y) \, F^n_{L/R} \left( x^\mu \right)} \ ,
\label{KK_S1/Z2_R__Func}
\end{equation}
where the 4D fields $F^n_{L/R}=Q^n_{L/R},D^n_{L/R}$ represent the KK states and satisfy the Dirac-Weyl equations,
\begin{equation}
\forall \, n \in \mathbb{N}, \ 
\left\{
\begin{array}{r c l}
i \bar \sigma^\mu \partial_\mu F_L^n \left( x^\mu \right) - m_n \, F_R^n \left( x^\mu \right) &=& 0 \, ,
\\ \vspace{-0.2cm} \\
i \sigma^\mu \partial_\mu F_R^n \left( x^\mu \right) - m_n \, F_L^n \left( x^\mu \right) &=& 0 \, ,
\end{array}
\right.
\label{4D_Dirac__Func}
\end{equation}
involving the KK mass eigenvalues $m_n$, while the only resulting profiles $f^n_{L/R}(y)=q^n_{L/R}(y)$, $d^n_{L/R}(y)$, included respectively into $F=Q,D$, are vanishing
over $y \in [0, \pi R]$. The opposite signs in front of each mass term of the bulk 
profile EOM induced by Eq.~\eqref{ELE_S1/Z2_R__Func_Free_Chiral}, with respect to the calculations of Ref.~\cite{Angelescu:2019viv},
just originate from a different sign convention for the $\Gamma^4$ matrix [see Eq.~\eqref{gamma_3}] 
and hence do not modify the (un)physical result of vanishing profiles.
Now let us study the profile solutions in the complementary region, $y \in [-\pi R^+, 0^-]$. 
Inserting the KK decomposition~\eqref{KK_S1/Z2_R__Func} into the first type of $\mathbb{Z}_2$ transformation~\eqref{Parity_S1/Z2_R_1}, 
one obtains the $\mathbb{Z}_2$ transformations directly on the $f^n_{L/R}(y)$ profiles ($\forall \, n \in \mathbb{N}$):
\begin{equation} 
Type \,\uppercase\expandafter{\romannumeral1}
\left\{
\begin{array}{c c c}
\displaystyle{ \sum_{n=0}^{+\infty} \left[q^n_{L(R)}(-y) \varmp   q^n_{L(R)}(y)\right] \, Q^n_{L(R)} \left( x^\mu \right)} 
= 0 \, \Rightarrow  \, q^n_{L(R)}(-y) = \varpm   q^n_{L(R)}(y)
\\ \vspace{-0.2cm} \\
\displaystyle{ \sum_{n=0}^{+\infty} \left[d^n_{L(R)}(-y) \varpm  d^n_{L(R)}(y)\right] \, D^n_{L(R)} \left( x^\mu \right)} 
= 0\, \Rightarrow \, d^n_{L(R)}(-y) = \varmp   d^n_{L(R)}(y) 
\end{array}
\right.
\label{Parity_Pro_S1/Z2_R_1-Free}
\end{equation}
where the implications come from the linear independence of mass eigenstates $F^n_{L/R} \left( x^\mu \right)$. Similarly, for the three other types of
$\mathbb{Z}_2$ transformations~\eqref{Parity_S1/Z2_R_2}-\eqref{Parity_S1/Z2_R_4}, we have the following profile parities:
\begin{eqnarray} 
\uppercase\expandafter{\romannumeral2}
\left\{
\begin{array}{c c c}
q^n_{L(R)}(-y) = \varmp   q^n_{L(R)}(y) 
\\ \vspace{-0.2cm} \\
d^n_{L(R)}(-y) = \varpm   d^n_{L(R)}(y) 
\end{array}
\right.
\uppercase\expandafter{\romannumeral3}
\left\{
\begin{array}{c c c}
q^n_{L(R)}(-y) = \varpm   q^n_{L(R)}(y) 
\\ \vspace{-0.2cm} \\
d^n_{L(R)}(-y) = \varpm   d^n_{L(R)}(y) 
\end{array}
\right.
\nonumber \\
\uppercase\expandafter{\romannumeral4}
\left\{
\begin{array}{c c c}
q^n_{L(R)}(-y) = \varmp   q^n_{L(R)}(y) 
\\ \vspace{-0.2cm} \\
d^n_{L(R)}(-y) = \varmp   d^n_{L(R)}(y) 
\end{array}
\right.
\label{Parity_Pro_S1/Z2_R_4}
\end{eqnarray}
Therefore, all the $f^n_{L/R}(y)$ profiles are systematically vanishing on the whole $\mathcal{S}^1 / \mathbb{Z}_2$ orbifold region, $y \in [-\pi R^+, 0^-] \cup [0, \pi R]$. 
Such profiles conflict with the two (for $L/R$) ortho-normalisation conditions over the full domain,
\begin{equation}
\forall \, n, m \in \mathbb{N}, \ \dfrac{1}{2 \pi R} \left(\int_{-\pi R^+}^{0^-} + \int_{0}^{ \pi R}\right) dy \ f^{n*}_{L/R}(y) \, f^m_{L/R}(y) = \delta_{nm} \, ,
\label{KK_Normalization_S1/Z2_R__Func_Free}
\end{equation}
originating from the condition of a canonical form for the 4D effective kinetic terms.
Hence the solutions for the fields obtained through this first method are not physically consistent.

\subsection{Introducing the EBC}
\label{S1/Z2_R_Free_EBC__Func}

In fact, one necessary ingredient was missing in the naive approach of Section~\ref{S1/Z2_R_Free_NBC__Func}. In order to identify it, we have to study 
the conserved fermion probability currents corresponding, via the Noether's theorem, to the global ${\rm U(1)_{Q}}$ and ${\rm U(1)_{D}}$ symmetries of the
action, $$S_{bulk} \, ,$$ involving the Lagrangian~\eqref{S_Psi_S1/Z2_R__Func}. The two independent global ${\rm U(1)_{Q,D}}$ transformations of the fields, 
letting $\mathcal{L}_{kin}$ invariant, act respectively as,
\begin{align}
Q \mapsto \text{e}^{i \alpha} Q \, , \ \bar{Q} \mapsto \text{e}^{-i \alpha} \bar{Q} \, , \, {\rm and,} \ 
D \mapsto \text{e}^{i \alpha'} D \, , \ \bar{D} \mapsto \text{e}^{-i \alpha'} \bar{D}  \, ,
\label{U1sym_S1/Z2_R__Func} 
\end{align}
where $\alpha, \alpha'$ ($\in \mathbb{R}$) are continuous constants entering for instance the infinitesimal field variations~\footnote{Different clear notations are used 
here for the infinitesimal field variations under specific transformations, $\underline{\delta} F$, and the above generic field variations in the variation calculus context of the 
least action principle, $\delta F$ [see typically Eq.~\eqref{Sp_group}].}: $$\underline{\delta} Q = i \alpha Q \, , \ \ \underline{\delta} \bar Q = - i \alpha \bar Q \, .$$
Choosing instead to consider a unique symmetry ($\alpha =  \alpha'$ for any field $F$) would correspond to a particular case only, among the general 
Lagrangian symmetry possibilities. Besides, this particular case would not provide the maximal information, since one symmetry would be associated to 
only one conserved probability current. We thus well consider, in this subsection, the transformations~\eqref{U1sym_S1/Z2_R__Func} [with both possibilities, 
$\alpha \neq \alpha'$ or $\alpha =  \alpha'$] and the two independent ${\rm U(1)_{Q,D}}$ symmetries. Based on these two symmetries, and the bulk EOM whose standard
structure appears in Eq.~\eqref{S_Psi_S1/Z2_R__Func_HVP}, the Noether's theorem predicts the local conservation relation, 
\begin{equation}
\partial_M j_F^M=0 \, , 
\label{eq:ConsCurrent}
\end{equation}
for the two probability currents,
\begin{equation}
j^M_Q = - \alpha \bar{Q} \; \Gamma^M Q \, , \  j^M_D = - \alpha' \bar{D} \; \Gamma^M D \, ,
\label{current_Q&D_S1/Z2_R__Func}
\end{equation}
as derived in details within the Appendix~B of Ref.~\cite{Angelescu:2019viv}. This relation holds over the whole $\mathcal{S}^1 / \mathbb{Z}_2$ orbifold domain,
$y \in [-\pi R^+, 0^-] \cup [0, \pi R]$, since the sole bulk terms in the action infinitesimal variation -- under ${\rm U(1)_{Q,D}}$ transformation -- must vanish for any integration 
sub-region included inside the entire integration domain of the action precisely defined for the model. The mathematical consistency of the 
condition~\eqref{eq:ConsCurrent} imposes necessarily continuous 5-current components over all the model space-time and in particular a continuous $j_F^4$ along
$y \in [-\pi R^+, 0^-] \cup [0, \pi R]$~\footnote{Notice that this condition is in agreement with Eq.~\eqref{eq:defFP} which guarantees continuous fields along 
$y \in [-\pi R^+, 0^-] \cup [0, \pi R]$.}. Furthermore, a jump of the form, $j_F^4\vert_{0^-} \neq j_F^4\vert_{0}$, would not determine any field at the fixed point and thus would not lead 
to vanishing variations in Eq.~\eqref{S_Psi_S1/Z2_R__Func_HVP_P} that would modify the BC~\eqref{BC_S1/Z2_R__Func_Free_NBC} inducing  
non-physical solutions. A similar argument applies at the other fixed point, $y=\pi R\equiv -\pi R$. Hence, one has to consider 
the remaining model possibility, $j_F^4\vert_{0^-} = j_F^4\vert_{0}$ 
and~\footnote{A change must occur at both fixed points to cure the problems of the solutions worked out in previous subsection.} 
$j_F^4\vert_{-\pi R^+} = j_F^4\vert_{\pi R}$, so that this current component is continuous over all the range, $y \in (-\pi R, \pi R  ]$.
In particular, we can now write,
\begin{equation}
j_F^4\vert_{0^-} = j_F^4\vert_{0} = j_F^4\vert_{0^+} \, .
\label{eq:current-cont}
\end{equation}
This obtained relation must be compared with the following one, coming directly from the $\mathbb{Z}_2$ transformations of type~\eqref{Parity_S1/Z2_R_1}-\eqref{Parity_S1/Z2_R_4}
and $\gamma^5$ properties,
\begin{align}
\left. j_F^4 \right|_{0^-} &= - \alpha^{(\prime)} \left. \bar{F} \; \Gamma^4 F \right |_{0^-} 
= - \alpha^{(\prime)} \left. \left(\pm   \gamma^5 F\right)^{\dagger} \gamma^0 \left[ - i \gamma^5 \right] \left(\pm   \gamma^5 F\right)  \right|_{0^+} \nonumber \\
&= \left. \alpha^{(\prime)} F^{\dagger} \gamma^0 \gamma^5 \left[ - i \gamma^5\right] \left(\gamma^5 F\right)  \right|_{0^+} 
= \alpha^{(\prime)} \left. \bar{F} \; \Gamma^4 F \right |_{0^+}  = - \left. j_F^4 \right|_{0^+}  \, . 
\label{Current_Parity_S1/Z2_R__Func_Free_EBC} 
\end{align}
The combination of Eq.~\eqref{eq:current-cont} and Eq.~\eqref{Current_Parity_S1/Z2_R__Func_Free_EBC} gives rise to a vanishing current component at the fixed point:
\begin{equation}
j_F^4\vert_{0^-} = j_F^4\vert_{0} = j_F^4\vert_{0^+} = 0 \, .
\nonumber%\label{eq:current-comb}
\end{equation}
Similar arguments regarding the second fixed point imply obviously that,
\begin{equation}
j_F^4\vert_{\pi R^-} = j_F^4\vert_{\pi R} = j_F^4\vert_{-\pi R^+} = 0 \, ,
\nonumber%\label{eq:current-comb}
\end{equation}
so that, using the generic chiral decomposition~\eqref{Sp_group-2comp-bis}, we get the following current conditions,
\begin{equation}
\left. j_F^4 \right|_{0, \pi R} = i \alpha^{(\prime)} \left. \left( F^\dagger_L F_R - F^\dagger_R F_L \right) \right|_{0, \pi R} = 0 \, ,
\label{Current_BC_S1/Z2_R__Func_Free_EBC} 
\end{equation}
leading to the minimal Boundary Conditions (BC),
\begin{equation}
\left\{
\begin{array}{r c l}
F_L|_{0} &=& 0 \, ,
\\ &\text{or} \\
F_R|_{0} &=& 0 \, ,
\end{array}
\right.
and \,  \ 
\left\{
\begin{array}{r c l}
F_L|_{\pi R} &=& 0 \, ,
\\ &\text{or}  \\
F_R|_{\pi R} &=& 0 \, .
\end{array}
\right. \ \ \ \ \ \ {\rm [EBC]}
\label{Gen_BC_S1/Z2_R__Func_Free_EBC} 
\end{equation}
These BC induce systematically
the vanishing of all the brane terms in the varied action obtained in Eq.~\eqref{S_Psi_BT_S1/Z2_R__Func_HVP_P}. Indeed, for example, the fixed value $F_L|_{0}=0$
implies $F^\dagger_L|_{0}=0$ and in turn $\delta F^\dagger_L|_{0}=0$~\footnote{Rigorously speaking, the action should 
not be minimised with respect to the known fixed fields so that the terms with vanishing field variations should not even appear. In fact, the brane terms of 
Eq.~\eqref{S_Psi_BT_S1/Z2_R__Func_HVP_P} should originally be written as a generic sum over unfixed fields.} 
[considering more precisely their two respective components as is clear from Appendix~\ref{app:Spin.1}]. Therefore the sole remaining BC are those of 
Eq.~\eqref{Gen_BC_S1/Z2_R__Func_Free_EBC}: there are no more NBC generated from the brane terms of Eq.~\eqref{S_Psi_BT_S1/Z2_R__Func_HVP_P}
and we name the BC~\eqref{Gen_BC_S1/Z2_R__Func_Free_EBC} as EBC since they are imposed by the $\mathbb{Z}_2$ 
transformations~\eqref{Current_Parity_S1/Z2_R__Func_Free_EBC} which contribute to define the studied model. From the point of view of the 
methodology, notice interestingly that it was necessary to consider the fermion probability currents to reveal the existence of the EBC. 
Now, solving the new EBC~\eqref{Gen_BC_S1/Z2_R__Func_Free_EBC} together with the unchanged bulk EOM~\eqref{ELE_S1/Z2_R__Func_Free_Chiral}  
over the domain, $y \in [0, \pi R]$, was precisely realised in Ref.~\cite{Angelescu:2019viv} where the same Eq.~(3.3) 
and (3.16) [there] were solved over the interval, $y \in [0, L]$. Once more, since the fields are continuous over $y \in [0, \pi R]$ [see Eq.~\eqref{eq:defFP}]
like there over $y \in [0, L]$, we can apply here the results derived in this previous work: the 5D solutions found for  
Eq.~\eqref{ELE_S1/Z2_R__Func_Free_Chiral}-\eqref{Gen_BC_S1/Z2_R__Func_Free_EBC} are given by 
Eq.~\eqref{KK_S1/Z2_R__Func}-\eqref{4D_Dirac__Func} and the following four possible sets of profiles over $y \in [0, \pi R]$ together with the associated KK mass spectrum 
equations ($\forall \, n \in \mathbb{N}$),
\begin{eqnarray}
1) & (--): \ f_{L}^n(y) = B^n_{L} \; \sin(m_n \, y) \, , \  (++): \ f_{R}^n(y) = B^n_{L} \; \cos(m_n \, y) \, ; \  \sin(m_n\, \pi R)=0 \, ,
\nonumber  \\ 
2) & (++): \ f_{L}^n(y) = B^n_{R} \; \cos(m_n \, y) \, , \  (--): \ f_{R}^n(y) = - B^n_{R} \;  \sin(m_n \, y) \, ; \  \sin(m_n\, \pi R)=0 \, ,
\nonumber  \\ \label{completeEBCsm0} 
\end{eqnarray}
and,
\begin{eqnarray}
3)  & (-+): \ f_{L}^n(y) = B^n_{L} \; \sin(m_n\,  y) \, , \  (+-): \ f_{R}^n(y) = B^n_{L} \; \cos(m_n \, y) \, ; \  \cos(m_n \,  \pi R)=0 \, ,
\nonumber  \\ 
4)  & (+-): \ f_{L}^n(y) = B^n_{R} \; \cos(m_n \,  y) \, , \  (-+): \ f_{R}^n(y) = - B^n_{R} \; \sin(m_n \, y) \, ; \  \cos(m_n\, \pi R)=0 \, .
\nonumber  \\ \label{completeEBCcusto0} 
\end{eqnarray}
The opposite signs in front of the $(--)$ and $(-+)$ profiles, with respect to the results in Ref.~\cite{Angelescu:2019viv},
just come from a different sign convention for the $\Gamma^4$ matrix, between here [see Eq.~\eqref{gamma_3}] and this reference.
In Eq.~\eqref{completeEBCsm0}-\eqref{completeEBCcusto0}, we use the standard BC notations, i.e. $-$ or $+$ for instance at $y=0$ stands 
respectively for the Dirichlet or Neumann BC: $f_{L/R}^n(0) = 0$ or $\partial_4 f_{L/R}^n(y)\vert_0 = 0$. For example, the symbolic notation $(-+)$ 
denotes Dirichlet (Neumann) BC at $y=0$ ($y=\pi R$). These notations make explicit the correspondence between the four EBC~\eqref{Gen_BC_S1/Z2_R__Func_Free_EBC}
and the four solutions~\eqref{completeEBCsm0}-\eqref{completeEBCcusto0}. The equation $\sin(m_n\,  \pi R)=0$ possesses the following solutions for the KK mass spectrum, 
\begin{eqnarray}
m_n=\pm \frac{n}{R} , \, n \in \mathbb{N} \, .
\label{Chiral_MS_S1/Z2_R__Func_Free_EBC} 
\end{eqnarray}
Similarly, the equation $\cos(m_n\,  \pi R)=0$ has the solutions: 
\begin{eqnarray}
m_n= \pm \frac{2n+1}{2R} , \, n \in \mathbb{N} \, .
\label{Cus_MS_S1/Z2_R__Func_Free_EBC} 
\end{eqnarray}
The part of the general $f_{L/R}^n(y)$ solutions in the complementary domain, $y\in [-\pi R^+, 0^-]$, 
is now obtained via the four types of $\mathbb{Z}_2$ transformations~\eqref{Parity_Pro_S1/Z2_R_1-Free}-\eqref{Parity_Pro_S1/Z2_R_4}.
Therefore, the inclusion of the EBC based on the vanishing probability currents allows to obtain consistent fermion profile and mass solutions.

\vspace{0.9cm}
\begin{table}[h!]
\centering
\resizebox{\textwidth}{!}
{
\begin{tabular}{| c | c | c | c | c | c | }
\hline
%\multirow{3}{*}{Continuity} 
 & \multirow{4}{*}{$\mathbb{Z}_2$} & \multicolumn{4}{ c |}{Fields} \\
\cline{3-6}
Continuity &  & \multicolumn{2}{ c |}{$Q_{L/R}$} & \multicolumn{2}{ c |}{$D_{L/R}$}  \\
\cline{3-6}
domains &  & \multirow{2}{*}{$q^n_L(y)/e^{i\alpha_Q^n}$} &  \multirow{2}{*}{$q^n_R(y)/e^{i\alpha_Q^n}$} 
 & \multirow{2}{*}{$d^n_L(y)/e^{i\alpha_D^n}$} &  \multirow{2}{*}{$d^n_R(y)/e^{i\alpha_D^n}$} \\
&  &  &  &  &  \\ 
\hline
$[0, \pi R]$ & Any & $\sqrt{2} \; \cos(m_n \; y)$ & $-\sqrt{2} \; \sin(m_n \; y)$ & $\sqrt{2} \; \sin(m_n \; y)$ & $\sqrt{2} \; \cos(m_n \; y)$ \\
\hline
\multirow{4}{*}{$[-\pi R^+, 0^-]$}
 & \uppercase\expandafter{\romannumeral1} & $\sqrt{2} \; \cos(m_n \; y)$ & $- \sqrt{2} \; \sin(m_n \; y)$ & $\sqrt{2} \; \sin(m_n \; y)$ & $\sqrt{2} \; \cos(m_n \; y)$ \\
\cline{2-6}
 & \uppercase\expandafter{\romannumeral2} & $- \sqrt{2} \; \cos(m_n \; y)$ & $\sqrt{2} \; \sin(m_n \; y)$ & $-\sqrt{2} \; \sin(m_n \; y)$ & $-\sqrt{2} \; \cos(m_n \; y)$ \\
\cline{2-6}
  & \uppercase\expandafter{\romannumeral3} & $\sqrt{2} \; \cos(m_n \; y)$ & $-\sqrt{2} \; \sin(m_n \; y)$ & $-\sqrt{2} \; \sin(m_n \; y)$ & $-\sqrt{2} \; \cos(m_n \; y)$ \\
\cline{2-6}
& \uppercase\expandafter{\romannumeral4} & $- \sqrt{2} \; \cos(m_n \; y)$ & $\sqrt{2} \; \sin(m_n \; y)$ & $\sqrt{2} \; \sin(m_n \; y)$ & $\sqrt{2} \; \cos(m_n \; y)$ \\
\hline\hline
{KK Masses} & \multicolumn{5}{ c |}{$|m_n| = n/R$, $n \in \mathbb{N}$}  \\
\hline
\end{tabular}
}
\vspace{0.25cm}
\caption{SM-like free fermionic $f^n_{L/R}(y)$ profiles -- normalised to the indicated complex phases -- on the two orbifold domains 
$[-\pi R^+, 0^-]$ and $[0, \pi R]$, corresponding to the solution of line~1 (2) in Eq.~\eqref{completeEBCsm0} for the field $D$ ($Q$).
The associated mass spectrum~\eqref{Chiral_MS_S1/Z2_R__Func_Free_EBC} is included as well for completeness. The profiles are given for the four types of $\mathbb{Z}_2$ 
transformations~\eqref{Parity_Pro_S1/Z2_R_1-Free}-\eqref{Parity_Pro_S1/Z2_R_4}. The phases $\alpha^n_{Q/D}$ belong to $\mathbb{R}$. 
In the special case, $n=0$, the $\sqrt{2}$ factors must all be replaced by the unity.}
\label{tab:SMQ&D_Free_Z2}
\end{table}
\vspace{0.25cm}

In Table~\ref{tab:SMQ&D_Free_Z2}, we present the explicit solutions over the whole orbifold domain for the SM-like profile $d_{L/R}^n(y)$ ($q_{L/R}^n(y)$)
taken from line~1 (2) of Eq.~\eqref{completeEBCsm0}: see the discussion on SM chirality configuration in Section~\ref{1_brTermSec}. 
The mass spectrum for the 4D KK states is defined by Eq.~\eqref{4D_Dirac__Func} and it is already determined by 
Eq.~\eqref{Chiral_MS_S1/Z2_R__Func_Free_EBC}-\eqref{Cus_MS_S1/Z2_R__Func_Free_EBC}. Notice on Table~\ref{tab:SMQ&D_Free_Z2} that the 
same $m_n$ spectrum enters the profile solutions in both regions, $y \in [0, \pi R]$, and, $y\in [-\pi R^+, 0^-]$. 
In this table, we also give the general values of the $B_{L/R}^n$ complex constants, in Eq.~\eqref{completeEBCsm0}, obtained from the ortho-normalisation 
conditions~\eqref{KK_Normalization_S1/Z2_R__Func_Free}~\footnote{Here, thanks to the profile parities,
a change of variable, $y\to-y$, could be applied to recover exclusively the integration domain $[0, \pi R]$.}.
We observe on Table~\ref{tab:SMQ&D_Free_Z2} that the choice of type of $\mathbb{Z}_2$ transformation is just a convention since it can modify the profile signs but it affects neither 
the mass spectrum nor the fermion chirality configuration -- as a certain chiral zero-mode profile vanishing on the region $[0, \pi R]$ is also systematically vanishing over $y\in [-\pi R^+, 0^-]$. 
In contrast, the chirality configuration and mass spectrum are fixed by the choice of EBC~\eqref{Gen_BC_S1/Z2_R__Func_Free_EBC} which can lead either to the two kinds of chiral 
solutions in Eq.~\eqref{completeEBCsm0} or to the vector-like solutions~\eqref{completeEBCcusto0}.

\begin{figure}[h]
\centering
\includegraphics[width=15cm]{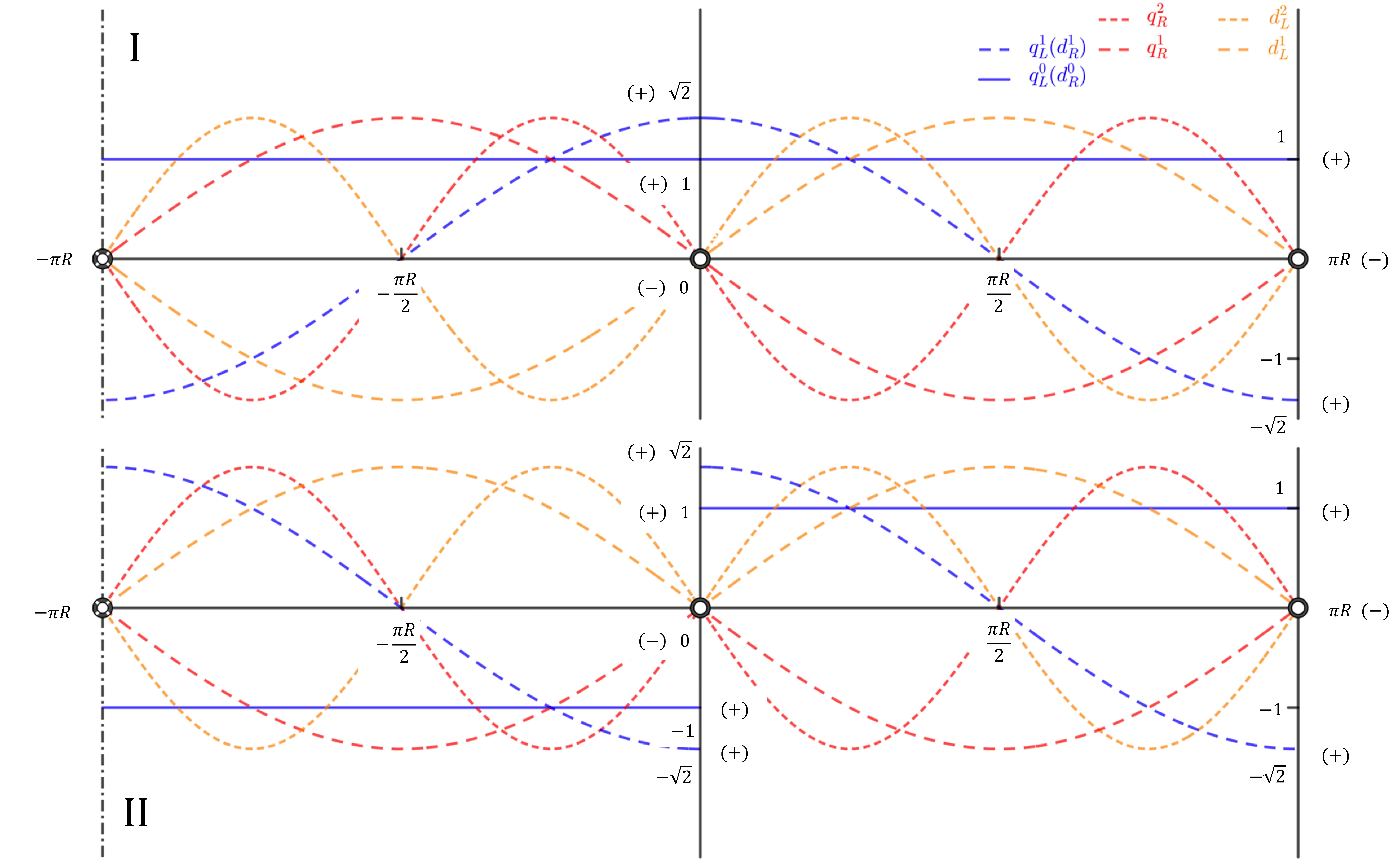}
\vspace{0.25cm}
\caption{Zero-mode and KK dimensionless
wave functions $q^n_{L/R}(y)$, $d^n_{L/R}(y)$, with $n=0,1,2$, along the $\mathcal{S}^1/ \mathbb{Z}_2$ orbifold domain, $y \in [-\pi R^+, 0^-] \cup [0, \pi R]$, 
corresponding to the free solutions of Table~\ref{tab:SMQ&D_Free_Z2} in the simplified case, $\alpha^n_{Q,D}=0$, $m_n>0$, and for the two different types of 
$\mathbb{Z}_2$ transformations, $\uppercase\expandafter{\romannumeral1}$, $\uppercase\expandafter{\romannumeral2}$ from 
Eq.~\eqref{Parity_Pro_S1/Z2_R_1-Free}-\eqref{Parity_Pro_S1/Z2_R_4}. The two fixed points at, $y=0$, $y=\pi R\equiv -\pi R$, and Dirichlet/Neumann BC, $(-)/(+)$, 
are indicated on the graph.}
\label{Profiles_S1/Z2_R_Free_Graphic}
\end{figure}

In Figure~\ref{Profiles_S1/Z2_R_Free_Graphic}, we draw the first two excitation profiles for each free solution presented in Table~\ref{tab:SMQ&D_Free_Z2} within the
simple real case, $\alpha^n_{Q,D}=0$, and for two different types of $\mathbb{Z}_2$ transformations from Eq.~\eqref{Parity_Pro_S1/Z2_R_1-Free}-\eqref{Parity_Pro_S1/Z2_R_4}. 
We see clearly on Figure~\ref{Profiles_S1/Z2_R_Free_Graphic} that for example with the type~$\uppercase\expandafter{\romannumeral2}$ of $\mathbb{Z}_2$ transformation, 
jumps appear for the profiles $q^{0,1}_{L}(y)$ and $d^{0,1}_{R}(y)$ at the two fixed points at, $y=0$, $y=\pi R\equiv -\pi R$, in the scenario without Yukawa couplings.
The presence of profile discontinuities here already justifies the treatment exposed in Section~\ref{5D_Geo}. The precise prescription~\eqref{eq:defFP} regarding the action integration 
domain, described in this section, renders the jumps of Figure~\ref{Profiles_S1/Z2_R_Free_Graphic} consistent mathematically: the difference, e.g. $q^{1}_{L}(0^-) \neq q^{1}_{L}(0)$, 
is compatible with a well defined Lagrangian integrand over the action integration domain, $y \in [-\pi R^+, 0^-] \cup [0, \pi R]$, where the profiles are continuous.

\vspace{1cm}

\subsection{Introducing the BBT}
\label{S1/Z2_R_Free_BBT__Func}

As suggested in Section~\ref{1_brTermSec}, we can alternatively introduce the dimension 4 operators of Eq.~\eqref{S_BBT_S1/Z2_R__Func} to study their effects
with respect to the inconsistencies raised in Section~\ref{S1/Z2_R_Free_NBC__Func}. Hence, to the action $S_{bulk}$ from Eq.~\eqref{eq:Sstart}, we add now another part 
and consider: $$S_{bulk} + S_B\, .$$
The variations of $S_B$ with respect to the generic field $\bar{F}$ [using Eq.~\eqref{Sp_group-2comp}],
\begin{equation}
\delta_{\bar{F}} S_{B} = \displaystyle{ \int d^4x \ \left(\sigma^F_{0} \left. \delta F^\dagger_L F_R \right|_{0} 
+ \sigma^F_{0} \left. \delta F^\dagger_R F_L \right|_{0} + \sigma^F_{\pi R} \left. \delta F^\dagger_L F_R \right|_{\pi R} 
+ \sigma^F_{\pi R} \left. \delta F^\dagger_R F_L \right|_{\pi R} \right)} \, ,
\nonumber%\label{S_BBT_S1/Z2_R__Func_HVP}
\end{equation}
together with Eq.~\eqref{S_Psi_BT_S1/Z2_R__Func_HVP_P} allow to write down the variations of the free fermion action:
\begin{align}
&\delta_{\bar{F}} \left( S_{bulk} + S_{B} \right) = \displaystyle{ \int d^4x \, \left\{ \left(\int_{-\pi R^+}^{0^-} + \int_{0}^{ \pi R}\right) dy \ \delta {\bar{F}} \, i \Gamma^{M} 
\partial_{M} F  \right.} + \left(\sigma^F_{0}1 + 1\right) \left. \delta F^\dagger_L F_R \right|_{0} 
\nonumber \\
&\displaystyle{\ \ \ \ \ \ + \left(\sigma^F_{0}1 - 1\right) \left. \delta F^\dagger_R F_L \right|_{0} } 
\displaystyle{\left. +\left(\sigma^F_{\pi R}1 - 1\right) \left. \delta F^\dagger_L F_R \right|_{\pi R} + \left(\sigma^F_{\pi R}1 + 1\right) \left. \delta F^\dagger_R F_L \right|_{\pi R}
\right\}} .
\label{S_Q_S1/Z2_R__Func_HVP_Free_BBT}
\end{align}
The individual vanishing of those volume and surface terms lead to the EOM~\eqref{ELE_S1/Z2_R__Func_Free_Chiral} together with the four following NBC,
depending on the two $\sigma^F_{0,\pi R}$ choices,
\begin{equation}
\left\{
\begin{array}{r c l}
F_L|_{0} &=& 0 \  (\sigma^F_{0} =-) ,
\\ &\text{or} \\
F_R|_{0} &=& 0 \  (\sigma^F_{0} =+) ,
\end{array}
\right.
and \,  \ 
\left\{
\begin{array}{r c l}
F_L|_{\pi R} &=& 0 \  (\sigma^F_{\pi R} =+) ,
\\ &\text{or}  \\
F_R|_{\pi R} &=& 0 \  (\sigma^F_{\pi R} =-) .
\end{array}
\right. \ \ \ \ \ \ {\rm [NBC]}
\label{Gen_BC_S1/Z2_R__Func_Free_BBT} 
\end{equation}
At this level, the EOM and NBC are effectively the same as the EOM~\eqref{ELE_S1/Z2_R__Func_Free_Chiral} and EBC~\eqref{Gen_BC_S1/Z2_R__Func_Free_EBC}
of the previous subsection, in the domain $y\in [0,\pi R]$, so that we find again the solutions~\eqref{completeEBCsm0}-\eqref{completeEBCcusto0} together with the
mass spectra~\eqref{Chiral_MS_S1/Z2_R__Func_Free_EBC}-\eqref{Cus_MS_S1/Z2_R__Func_Free_EBC}. For instance, the SM-like 
choice $\sigma^Q_{0(\pi R)} =\varpm$ of Eq.~\eqref{S_BBT_S1/Z2_R__Func} leads via Eq.~\eqref{Gen_BC_S1/Z2_R__Func_Free_BBT} 
to the solution of line~2 in Eq.~\eqref{completeEBCsm0}. Then the parts of the general profile solutions in the complementary region, $y\in [-\pi R^+, 0^-]$, 
are found out via the different types of $\mathbb{Z}_2$ transformations~\eqref{Parity_Pro_S1/Z2_R_1-Free}-\eqref{Parity_Pro_S1/Z2_R_4} in the free case, as in 
Section~\ref{S1/Z2_R_Free_EBC__Func}, so that the complete solutions are once more identical and can also be illustrated by the Table~\ref{tab:SMQ&D_Free_Z2}
and Figure~\ref{Profiles_S1/Z2_R_Free_Graphic} both based on the ortho-normalisation conditions~\eqref{KK_Normalization_S1/Z2_R__Func_Free}. 
In conclusion, introducing the BBT permits to rigorously work out profile and mass solutions. A second conclusion in this approach is that the chirality set-up  
-- one of the two chiral solutions~\eqref{completeEBCsm0} or of the vector-like ones~\eqref{completeEBCcusto0} -- and associated mass spectrum 
are fixed by the choice of NBC~\eqref{Gen_BC_S1/Z2_R__Func_Free_BBT} and thus originally by the choices of $\sigma^F_{0,\pi R}$ BBT signs in 
Eq.~\eqref{S_BBT_S1/Z2_R__Func}. In simpler words, the BBT (like the EBC previously) control the chiral nature of the theory as well as each field chirality.

Let us now discuss the probability currents. The addition of the $S_B$ part in Eq.~\eqref{S_BBT_S1/Z2_R__Func} to $S_{bulk}$ is not affecting 
the current equations~\eqref{eq:ConsCurrent}-\eqref{current_Q&D_S1/Z2_R__Func} since the new brane terms so induced in the infinitesimal action variation 
-- under the ${\rm U(1)_{Q,D}}$ transformations~\eqref{U1sym_S1/Z2_R__Func} -- vanish due to their ${\rm U(1)_{Q,D}}$ invariant form. In contrast with the previous subsection
and with the interval model in the free case with BBT~\cite{Angelescu:2019viv}, there exists no demonstration here of Eq.~\eqref{Current_BC_S1/Z2_R__Func_Free_EBC}. 
Nevertheless, we can check that $j_F^4\vert_{0,\pi R}$ is well vanishing by using the obtained solutions~\eqref{completeEBCsm0}-\eqref{completeEBCcusto0}: 
the product $f_{L}^n(y) f_{R}^m(y)$ systematically vanishes at $y=0,\pi R$. Therefore, the BBT play the r\^ole of making $j_F^4\vert_{0,\pi R}$ vanish ($\mathbb{Z}_2$
transformation consequence) like the EBC were guaranteeing it in Section~\ref{S1/Z2_R_Free_EBC__Func}. Note that we could simultaneously apply 
the EBC and introduce the BBT but those two processes would be physically redundant to define the model.

\section{Brane-localised scalar couplings in the orbifold: 4D approach}
\label{S1/Z2_R_Yuk_4D__Func}

Once the free case is addressed, via the EBC~\eqref{Gen_BC_S1/Z2_R__Func_Free_EBC}  in Section~\ref{S1/Z2_R_Free_EBC__Func} 
or the NBC~\eqref{Gen_BC_S1/Z2_R__Func_Free_BBT} induced by the BBT in Section~\ref{S1/Z2_R_Free_BBT__Func}, 
the free fermion mass spectrum and profiles are known. Then how to take into account the effects of the action part $S_{X}$ in the mass spectrum,
the action~\eqref{S_X_S1/Z2_R__Func} being induced by the Yukawa interaction between a brane-localised scalar field and bulk fermions? 
The considered action reads thus as,
\begin{align} 
S_{bulk} + S_{X} \ (+ S_{B}) \ .
\label{S_S1/Z2_R__Func_VEV-4Dap}
\end{align}
A first method called the perturbation method, 
described in the present section, is performed at the level of the 4D effective Lagrangian, that is by calculating the mass mixings between the 
different levels of the KK towers. Considering the SM-like profile solutions $d_{L/R}^n(y)$ ($q_{L/R}^n(y)$) and associated free KK mass spectrum from line~1 (2) of 
Eq.~\eqref{completeEBCsm0}, all the initial 4D effective masses for the KK modes of Eq.~\eqref{KK_S1/Z2_R__Func} in the interaction basis can be classified into 
two species: the pure KK masses~\eqref{Chiral_MS_S1/Z2_R__Func_Free_EBC} and the mass contributions from the Yukawa interaction
given by the overlap between the wave functions and Higgs-brane,
\begin{equation}
\left\{
\begin{array}{l}
\displaystyle{\forall (i, j) \in \mathbb{N}^2, \, \alpha_{ij} = X \, \dfrac{q_L^i(\pi R)}{\sqrt{2\pi R}} \, \dfrac{d_R^j(\pi R)}{\sqrt{2\pi R}} } \, , \\ \\ \vspace{-0.8cm} \\
\displaystyle{\forall (i, j) \in \mathbb{N}^{\star 2}, \, \beta_{ij} = X' \, \dfrac{d_L^i(\pi R)}{\sqrt{2\pi R}} \, \dfrac{q_R^j(\pi R)}{\sqrt{2\pi R}} } \, .
\end{array}
\right.
\label{Coeff_S1/Z2_R_4D__Func}
\end{equation}
In particular, $\beta_{ij}=0$ as imply the respective SM solutions~\eqref{completeEBCsm0} so that the coupling constant $X'$ disappears from 
the mass dependences. Note that for similar reasons [{\it c.f.} Eq.~\eqref{Sp_group-2comp-bis}], 
in case of the presence of the BBT~\eqref{S_BBT_S1/Z2_R__Func}, those do not generate 4D mass terms.
All the 4D mass terms enter the 4D effective Lagrangian through the following mass matrix, 
\begin{equation}
- \chi^\dagger_L \, \mathcal{M} \, \chi_R + \text{H.c.}
\nonumber%\label{Yuk_S1/Z2_R_Yuk_4D__Func}
\end{equation}
within the field basis noted,
\begin{equation}
\left\{
\begin{array}{r c l}
\chi^t_L (x^\mu) &=& \left( Q_L^{0t}, Q_L^{1t}, D_L^{1t}, Q_L^{2t}, D_L^{2t}, \cdots \right), \\ \\
\chi^t_R (x^\mu) &=& \left( D_R^{0t}, Q_R^{1t}, D_R^{1t}, Q_R^{2t}, D_R^{2t}, \cdots \right) .
\label{Chi_S1/Z2_R_Yuk_4D__Func}
\end{array}
\right.
\end{equation}
The texture of this infinite mass matrix $\mathcal{M}$ involving the diagonal  
$m_n$, off-diagonal $\alpha_{ij}$ and mixing the $Q$, $D$ fields together can be precisely taken from the 
interval model context~\cite{Angelescu:2019viv} (Section~3.2), with the replacement $L \leftrightarrow \pi R$, since the KK masses and bulk profile solutions are 
then identical (up to extensions over $[-\pi R^+,0^-]$ as seen in Section~\ref{S1/Z2_R_Free_EBC__Func} here) like the Yukawa interactions localised at $y=\pi R$
[for any $\mathbb{Z}_2$ transformation~\eqref{Parity_Pro_S1/Z2_R_1-Free}-\eqref{Parity_Pro_S1/Z2_R_4}]. 
Now we can apply the results for the mass eigenvalues $M_n$ of the 4D eigenstates $\psi^n_{L/R}(x^\mu)$ obtained through the bi-diagonalisation performed 
in this Ref.~\cite{Angelescu:2019viv}, based on the calculations of 
Ref.~\cite{Barcelo:2014kha}, by re-normalising $X$ to $X/2$ since the two present profiles (even or odd) entering 
$\alpha_{ij}$ are normalised via Eq.~\eqref{KK_Normalization_S1/Z2_R__Func_Free}
over a domain of double size $2L \leftrightarrow 2\pi R$ compared to the interval case. Doing so, the obtained exact mass eigenvalues are determined by the
following equation, coming from the characteristic equation,
\begin{equation}
\forall \, n \in \mathbb{N}, \ \tan^2(\sqrt{|M_n|^2} \, \pi R) = \left(\dfrac{X}{2}\right)^2  ,
\label{MS_S1/Z2_R__Func_Yuk_4D_Origin}
\end{equation}
in the case of a real $X$ parameter
and the positive $m_n$ branch from Eq.~\eqref{Chiral_MS_S1/Z2_R__Func_Free_EBC}.
Notice that the different conventional sign in front of the $(--)$ profiles found in Eq.~\eqref{completeEBCsm0} [here $q^n_R(y)$ and $d^n_L(y)$], 
with respect to the interval study~\cite{Angelescu:2019viv},
does not affect the final mass spectrum -- as is clear from Eq.~\eqref{Coeff_S1/Z2_R_4D__Func}. 
Hence, the physical absolute value of the mass spectrum reads as:
\begin{equation}
|M_n| = \dfrac{1}{\pi R} \left \vert \arctan \left( \dfrac{X}{2} \right) + (-1)^n \, \tilde n(n) \, \pi \right \vert , \, n \in \mathbb{N} \, ,
\label{MS_S1/Z2_R__Func_Yuk_4D}
\end{equation}
with the function $\tilde n(n)$ defined according to,
\begin{equation}
\tilde n(n) =
\left\{
\begin{array}{r c l}
& \frac{n}{2} \ {\rm for} \ n \ {\rm even}\, ,
\\ \vspace{-0.2cm} \\
& \frac{n+1}{2} \ {\rm for} \ n \ {\rm odd}\, .
\end{array}
\right. 
\label{ntilde}
\end{equation}

% ((
% Mass matrix: BBT used for free mass and reput ((ok but not in general to get something meaningful)) for 4D effective (4D method) as not used yet for Yuk
% ((not like in 5D method: BBT used for Yuk so not reput for 4D effective Yuk 
%                                                   CONFIRMED BY MATCHING but subtle as Yuk reput as part of Lag not for spatial properties yet taken))

% Mass matrix (and calculation principle): use interval paper without regul
% as same field content, matrix structure ((m_n, m_Y)), KK mass form L=piR (duality appears as see in Section ) and Yukawa mass form
% since respectively same bulk fermions and brane interaction type 

% m_n alpha/beta expressions up to factor(s) as different normalisation as different size for L=piR: allows to apply...
% Mass eigenvalues: use Barcelo's paper for calculation with KK mass replacement L=piR  
% ))

\section{Brane-localised scalar couplings in the orbifold: 5D approach}
\label{S1/Z2_R_Yuk__Func}

\subsection{Applying the NBC}
\label{S1/Z2_R_Yuk_NBC__Func}
 
Let us now study the presence of Yukawa couplings at the fixed point, $y=\pi R$, through the action,
\begin{align}
S_{bulk} + S_{X} + S_B^0 \, , \ {\rm with,} \ S_B^0 =  \int d^4x \ \left ( \sigma^Q_{0} \, \left. \bar{Q}Q \right|_{0} +  \sigma^D_{0} \, \left. \bar{D}D \right|_{0} \right )\, , 
\label{S_S1/Z2_R__Func_VEV-4Dap-3}
\end{align}
within the 5D approach, that is by considering the mixings among KK excitation states at the level of the 5D fields.
The BBT introduced here at the fixed point at $y=0$ are the ones of Eq.~\eqref{S_BBT_S1/Z2_R__Func} leading to SM-like chirality configurations:
$\sigma^Q_{0} =+$, $\sigma^D_{0} = -$. Those guarantee a correct treatment of the free brane, like the EBC, as analysed throughout Section~\ref{S1/Z2_R_Free__Func}.
Using Eq.~\eqref{S_Q_S1/Z2_R__Func_HVP_Free_BBT} and Eq.~\eqref{S_X_S1/Z2_R__Func}, one gets directly the following action variations with respect to the fields 
$\bar{Q}$ and $\bar{D}$, 
\begin{align}
\delta_{\bar{Q}} (S_{bulk} + S_{X} + S_B^0) &=\displaystyle{ \int d^4x \, \left\{ \left(\int_{-\pi R^+}^{0^-} + \int_{0}^{ \pi R}\right) dy 
\ \delta {\bar{Q}} \, i \Gamma^{M} \partial_{M} Q  \right.} 
\label{S_D_S1/Z2_R__Func_Yuk_HVP_EBC-Yuk} 
+ \\ 
&\displaystyle{\left. \left. \left[\delta Q^\dagger_L \left(- XD_R - Q_R  \right) + \delta Q^\dagger_R \left(- X'D_L + Q_L \right) \right] \right|_{\pi R} 
+ 2 \left(\left. \delta {Q^\dagger_L}  Q_R \right)\right|_{0} \right\}} \,  , 
\nonumber \\
\delta_{\bar{D}} (S_{bulk} + S_{X} + S_B^0) &=\displaystyle{ \int d^4x \, \left\{ \left(\int_{-\pi R^+}^{0^-} + \int_{0}^{ \pi R}\right) dy 
\ \delta {\bar{D}} \, i \Gamma^{M} \partial_{M} D \right.} + \nonumber \\ 
&\displaystyle{\left.  \left. \left[\delta D^\dagger_L \left( - X'^*Q_R - D_R \right) + \delta D^\dagger_R \left(- X^*Q_L + D_L \right)  \right] \right|_{\pi R} - 2 \left. \left(\delta {D^\dagger_R}  D_L  \right)\right|_{0} \right\}} \,  .
\nonumber
\end{align}
The separate vanishings of these volume and surface terms, induced by the least action principle, 
give rise respectively to the EOM~\eqref{ELE_S1/Z2_R__Func_Free_Chiral} and the following NBC,
\begin{equation}
\left\{
\begin{array}{l}
\left. \left( Q_R + X \; D_R \right) \right|_{\pi R} = 0 \, ,  \ \ \ \left. \left( D_L - X^{*}  \; Q_L \right) \right|_{\pi R} = 0 \, , \\
\left. \left( Q_L - X' \; D_L \right) \right|_{\pi R} = 0 \, ,  \ \ \ \left. \left( D_R + X^{\prime *}  \; Q_R \right) \right|_{\pi R} = 0 \, , \\
\left. Q_{R} \right|_0 = 0 \, , \ \left. D_{L} \right|_0 = 0 \, .
\end{array}
\right. 
\label{BC_S1/Z2_R__Func_Yuk_NBC}
\end{equation}
As usual, the 5D field solutions of the EOM~\eqref{ELE_S1/Z2_R__Func_Free_Chiral} and NBC~\eqref{BC_S1/Z2_R__Func_Yuk_NBC} have the form
of the following mixed KK decomposition [instead of Eq.~\eqref{KK_S1/Z2_R__Func}]~\cite{Casagrande:2008hr,Azatov:2009na},
\begin{equation}
\left [ 
\begin{array}{r c l}
Q_L \left( x^\mu, y \right) &=& \dfrac{1}{\sqrt{2 \pi R}} \displaystyle{ \sum_{n=0}^{+\infty} q^n_L(y) \, \psi^n_L \left( x^\mu \right) } \, ,
\\ \vspace{-0.3cm} \\
Q_R \left( x^\mu, y \right) &=& \dfrac{1}{\sqrt{2 \pi R}} \displaystyle{ \sum_{n=0}^{+\infty} q^n_R(y) \, \psi^n_R \left( x^\mu \right) } \, ,
\\ \vspace{-0.3cm} \\
D_L \left( x^\mu, y \right) &=& \dfrac{1}{\sqrt{2 \pi R}} \displaystyle{ \sum_{n=0}^{+\infty} d^n_L(y) \, \psi^n_L \left( x^\mu \right) } \, ,
\\ \vspace{-0.3cm} \\
D_R \left( x^\mu, y \right) &=& \dfrac{1}{\sqrt{2 \pi R}} \displaystyle{ \sum_{n=0}^{+\infty} d^n_R(y) \, \psi^n_R \left( x^\mu \right) } \, ,
\end{array} 
\right. 
\label{Mix_KK_S1/Z2_R__Func}
\end{equation}
with the 4D fields $\psi^n_{L/R}( x^\mu)$, already mentioned in Section~\ref{S1/Z2_R_Yuk_4D__Func}, satisfying the Dirac-Weyl equations,
\begin{equation}
\left\{
\begin{array}{r c l}
i \bar \sigma^\mu \partial_\mu \psi^n_L \left( x^\mu \right) - M_n \, \psi^n_R \left( x^\mu \right) &=& 0 \, ,
\\ \vspace{-0.2cm} \\
i \sigma^\mu \partial_\mu \psi^n_R \left( x^\mu \right) - M_n \, \psi^n_L \left( x^\mu \right) &=& 0 \, ,
\end{array}
\right.
\label{4D_Dirac_Mix__Func-Yuk}
\end{equation}
the $M_n$ being the fermion mass eigenvalues including the contributions from the Yukawa terms and
these 4D fields the mass eigenstates including the effects of mixings among the $Q$, $D$ fields as well as (infinite) KK levels.
The explicit profile solutions appearing in Eq.~\eqref{Mix_KK_S1/Z2_R__Func}
over the domain, $y \in [0, \pi R]$, were found out for the interval model studied in Ref.~\cite{Angelescu:2019viv} where the exactly identical EOM and same NBC,
up to a sign and a factor $2$ in front of each $X^{(\prime)}$ parameter,
[respectively Eq.~(3.3) and (5.5) there] have been solved over $y \in [0, L]$. Because the fields are continuous over $y \in [0, \pi R]$ [{\it c.f.} Eq.~\eqref{eq:defFP}]
like there over $y \in [0, L]$, one can apply here the conclusions obtained in this reference. The opposite sign in front of the $X^{(\prime)}$ parameters
originates from a different Dirac matrix sign convention [see the $\Gamma^4$ sign in Eq.~\eqref{gamma_3}] and has thus no physical consequences.
The relative factors $2$, at the same places in the NBC~\eqref{BC_S1/Z2_R__Func_Yuk_NBC}, come from the existence of surface terms both at 
$y=0$, $0^-$ and $y=\pi R$, $-\pi R^+$ as is clearly described in Eq.~\eqref{S_Psi_S1/Z2_R__Func_HVP_P}-\eqref{S_Psi_BT_S1/Z2_R__Func_HVP_P}.
These factors turn out not to modify the relations between the different profile solutions and to only induce a factor-$4$ change in the final mass spectrum equation, 
both obtained in Ref.~\cite{Angelescu:2019viv}. Besides, the two (for $L/R$) following ortho-normalisation conditions over the full $\mathcal{S}^1$ domain
[replacing Eq.~\eqref{KK_Normalization_S1/Z2_R__Func_Free}],
\begin{equation}
\forall \, n, m \in \mathbb{N}, \ \dfrac{1}{2 \pi R} \left(\int_{-\pi R^+}^{0^-} + \int_{0}^{ \pi R}\right) dy \, 
\left[ q^{n*}_{L/R}(y) \, q^m_{L/R}(y) + d^{n*}_{L/R}(y) \, d^m_{L/R}(y) \right] = \delta_{nm} \, ,
\label{Mix_KK_Normalization_S1/Z2_R__Func_Yuk}
\end{equation}
as induced by the decomposition~\eqref{Mix_KK_S1/Z2_R__Func}, can be recast into the integration relations 
of Ref.~\cite{Angelescu:2019viv} over the region $[0, \pi R]$ but with a global factor $2$,
thanks to the change of variable $y'=-y$, the fixed odd/even parities of the profiles and Eq.~\eqref{eq:defFP}, so that the demonstration about profile solutions 
on the interval in Ref.~\cite{Angelescu:2019viv} remains unchanged here, from this point of view as well. Indeed, 
injecting the mixed KK decomposition~\eqref{Mix_KK_S1/Z2_R__Func} into the first type of $\mathbb{Z}_2$ transformation~\eqref{Parity_S1/Z2_R_1}, 
we get the $\mathbb{Z}_2$ transformations directly on the profiles:
\begin{equation} 
Type \,\uppercase\expandafter{\romannumeral1}
\left\{
\begin{array}{c c c}
\displaystyle{ \sum_{n=0}^{+\infty} \left[q^n_{L(R)}(-y) \varmp   q^n_{L(R)}(y)\right] \, \psi^n_{L(R)} \left( x^\mu \right)} 
= 0 \, \Rightarrow  \, q^n_{L(R)}(-y) = \varpm   q^n_{L(R)}(y)
\\ \vspace{-0.2cm} \\
\displaystyle{ \sum_{n=0}^{+\infty} \left[d^n_{L(R)}(-y) \varpm  d^n_{L(R)}(y)\right] \, \psi^n_{L(R)} \left( x^\mu \right)} 
= 0\, \Rightarrow \, d^n_{L(R)}(-y) = \varmp   d^n_{L(R)}(y) 
\end{array}
\right.
\label{Parity_Pro_S1/Z2_R_1}
\end{equation}
In the same way, for the three other types of $\mathbb{Z}_2$ transformations~\eqref{Parity_S1/Z2_R_2}-\eqref{Parity_S1/Z2_R_4}, one 
obtains the same profile parities as in Eq.~\eqref{Parity_Pro_S1/Z2_R_4}. 
As a conclusion, the same result as in Ref.~\cite{Angelescu:2019viv} holds here for the orbifold: the 4D effective Yukawa coupling constant for the lightest modes
($\psi^0_{L,R}$), induced by the found profiles, tends to zero within the decoupling limit which is not compatible with the SM configuration expected.
The problematic characteristics of the solutions obtained in this naive approach are confirmed by the final mass spectrum equation, $\tan^2(M_n \, \pi R) = |X|^2$
(independent from the profile normalisations), which conflicts analytically with the one obtained through the 4D method in Eq.~\eqref{MS_S1/Z2_R__Func_Yuk_4D_Origin}
for a real $X$ parameter. This failure motivates the alternative 5D methods of the next two subsections.

\subsection{Introducing the EBC}
\label{S1/Z2_R_Yuk_EBC2__Func}
Following the same idea as for the free case in Section~\ref{S1/Z2_R_Free_EBC__Func}, we try now to find consistent fermion mass solutions via considerations 
on their currents. The currents permit a priori to fully define the geometrical field configuration like here for the $\mathcal{S}^1/ \mathbb{Z}_2$ orbifold scenario. 
The complete relevant action including the brane-localised Yukawa terms~\eqref{S_X_S1/Z2_R__Func}, 
\begin{align}
S_{bulk} + S_{X} + S_B^0 \, , 
\label{eq:Yuk-EBC-action}
\end{align}
like in Eq.~\eqref{S_S1/Z2_R__Func_VEV-4Dap-3}, is invariant under the unique ${\rm U(1)_F}$ symmetry defined via Eq.~\eqref{U1sym_S1/Z2_R__Func}
only for, 
\begin{equation}
\alpha = \alpha' \, , 
\label{AlphaEqual}
\end{equation}
since the fermions $Q$ and $D$ are mixed on the brane at $y=\pi R$. Based on this symmetry involving both $Q$ and $D$
as well as on the bulk EOM [whose standard
structure appears in the action variation~\eqref{S_Psi_S1/Z2_R__Func_HVP}], the Noether's theorem predicts ({\it c.f.} Appendix~B of Ref.~\cite{Angelescu:2019viv})
the new local probability conservation relation, 
\begin{equation}
\partial_M j^M = 0 \, ,  \phantom{0} \text{with,} \phantom{0} j^M = \displaystyle{\sum_{F=Q,D} j_F^M} \, ,
\label{current_Q+D_S1/Z2_R__Func}
\end{equation}
involving the individual currents given by Eq.~\eqref{current_Q&D_S1/Z2_R__Func}-\eqref{AlphaEqual} over the full orbifold domain, $y \in [-\pi R^+, 0^-]$     
$\cup [0, \pi R]$. Notice that 
the new $S_X$ brane terms entering the infinitesimal action variation -- under the ${\rm U(1)_F}$ transformations -- vanish because of their invariant form
and have thus no direct effect on the conservation relation~\eqref{current_Q+D_S1/Z2_R__Func}. The mathematical consistency of the 
relation~\eqref{current_Q+D_S1/Z2_R__Func} implies necessarily the continuity of 5-current components over the whole space-time and in particular a continuous 
$j^4$ along $y \in [-\pi R^+, 0^-] \cup [0, \pi R]$. Besides, a discontinuity of the form, $j^4\vert_{-\pi R^+} \neq j^4\vert_{-\pi R} \equiv j^4\vert_{\pi R}$, 
would not fix any field at this fixed point and in turn would not induce vanishing variations in Eq.~\eqref{S_D_S1/Z2_R__Func_Yuk_HVP_EBC-Yuk} possibly modifying 
the BC~\eqref{BC_S1/Z2_R__Func_Yuk_NBC} which induce the drawbacks already pointed out in Section~\ref{S1/Z2_R_Yuk_NBC__Func}. 
As a consequence, we must consider the remaining model possibility:
\begin{equation}
j^4\vert_{-\pi R^+} = j^4\vert_{-\pi R} \equiv j^4\vert_{\pi R} = j^4\vert_{\pi R^-} \, ,
\label{eq:current-cont-Yuk}
\end{equation}
where Eq.~\eqref{eq:defFP} is also invoked. On the other side, the current $j^4$ is odd under any type of $\mathbb{Z}_2$ 
transformation~\eqref{Parity_S1/Z2_R_1}-\eqref{Parity_S1/Z2_R_4} as can be shown in a similar way as in
Eq.~\eqref{Current_Parity_S1/Z2_R__Func_Free_EBC}:
\begin{align}
\left. j^4 \right|_{-\pi R^+} = - \left. j^4 \right|_{\pi R^-}  \, . 
\label{Current_Parity_S1/Z2_R__Func_Free_EBC-Yuk} 
\end{align}
Combining Eq.~\eqref{eq:current-cont-Yuk} with Eq.~\eqref{Current_Parity_S1/Z2_R__Func_Free_EBC-Yuk} leads to,
\begin{equation}
j^4\vert_{\pi R^-} = j^4\vert_{\pi R} = j^4\vert_{-\pi R^+} = 0 \, ,
\nonumber%\label{eq:current-comb}
\end{equation}
so that, using Eq.~\eqref{Current_BC_S1/Z2_R__Func_Free_EBC} and \eqref{AlphaEqual}, we get the relation (inducing EBC),
\begin{equation}
\left. j^4 \right |_{\pi R} = i \alpha \left. \left( Q^\dagger_L Q_R -  Q^\dagger_R Q_L + D^\dagger_L D_R - D^\dagger_R D_L \right) \right|_{\pi R} = 0 \, ,
\label{Current_BC_S1/Z2_R__Func_Yuk_EBC-Yuk} 
\end{equation}
and its variation (for a non-trivial transformation with $\alpha \neq 0$),
\begin{eqnarray}
\left( \delta Q^\dagger_L Q_R + Q^\dagger_L \delta Q_R - \delta Q^\dagger_R Q_L - Q^\dagger_R \delta Q_L  \right .
 \ \ \ \ \ \ \ \ \ \ \ \ \ \ \ \ \ \ \ \ \ \ \ \ \ \ \ \ \ \ \ \ \ \label{current_var_condition-Yuk} \\
\left . \left . + \, \delta D^\dagger_L D_R + D^\dagger_L \delta D_R  - \delta D^\dagger_R D_L - D^\dagger_R \delta D_L \right) \right|_{\pi R} = 0 \, .
\nonumber
\end{eqnarray}
At this level, we can consider the search for field solutions of vanishing Eq.~\eqref{S_D_S1/Z2_R__Func_Yuk_HVP_EBC-Yuk} 
and Eq.~\eqref{Current_BC_S1/Z2_R__Func_Yuk_EBC-Yuk}-\eqref{current_var_condition-Yuk} 
first on the domain, $y \in [0, \pi R]$, which is equivalent to the search performed for the interval model in 
Ref.~\cite{Angelescu:2019viv} with the replacement, $L \leftrightarrow \pi R$. Given that the ortho-normalisation 
condition~\eqref{Mix_KK_Normalization_S1/Z2_R__Func_Yuk} written on the domain $[0, \pi R]$ is the same within the 
orbifold and interval frameworks, up to an overall factor $2$, we can apply the conclusion of Ref.~\cite{Angelescu:2019viv} and claim that there
exists no SM-like consistent solution for the fields (over $y \in [0, \pi R]$) for similar reasons as in Section~\ref{S1/Z2_R_Yuk_NBC__Func}.
As a conclusion, the introduction of EBC does not constitute the correct approach towards the treatment of point-like Yukawa interactions at 
a fixed point of the $\mathcal{S}^1/ \mathbb{Z}_2$ orbifold. Regarding the bulk fermion probability currents, both the cases of a $j^4$ jump 
and a $j^4$ continuity at the Yukawa coupling location, $y=\pi R$, lead to inconsistent field solutions so that, at this stage of the study, there exists 
no theoretical proof of the $j^4$ continuity -- and via Eq.~\eqref{Current_Parity_S1/Z2_R__Func_Free_EBC-Yuk} of its vanishing -- at this fixed point, 
in contrast with the interval model (case of presence of boundary-localised Yukawa interactions)~\cite{Angelescu:2019viv}.

\subsection{Introducing the BBT}
\label{S1/Z2_R_Yuk_BBT__Func}

In order to get meaningful field solutions in the presence of brane-localised Yukawa couplings at the fixed point, $y=\pi R$, let us finally try the introduction of 
the SM-like BBT~\eqref{S_BBT_S1/Z2_R__Func} as in the free case of Section~\ref{S1/Z2_R_Free_BBT__Func} or as in the interval model~\cite{Angelescu:2019viv}.
We thus consider here the same action as in Eq.~\eqref{S_S1/Z2_R__Func_VEV-4Dap-3}-\eqref{eq:Yuk-EBC-action} but adding now the BBT at  $y=\pi R$:
\begin{align}
S_{bulk} + S_{X} + S_B \, . 
\nonumber%\label{eq:Yuk-BBT-action}
\end{align}
Using Eq.~\eqref{S_Q_S1/Z2_R__Func_HVP_Free_BBT} and Eq.~\eqref{S_D_S1/Z2_R__Func_Yuk_HVP_EBC-Yuk}, we find 
the following action variations with respect to $\bar{Q}$ and $\bar{D}$:
\begin{align}
\delta_{\bar{Q}} ( S_{bulk} + S_{X} + S_B ) &=\displaystyle{ \int d^4x \, \left\{ \left(\int_{-\pi R^+}^{0^-} + \int_{0}^{ \pi R}\right) dy 
\ \delta {\bar{Q}} \, i \Gamma^{M} \partial_{M} Q \right.}  
\nonumber%\label{eq:Y-orb_BBT-QD} 
\\ & \ \ \ + \displaystyle{\left. \left. \left[-2\, \delta Q^\dagger_L \left(Q_R + \dfrac{X}{2} D_R \right) - X' \delta Q^\dagger_R D_L  \right] \right|_{\pi R} 
+ 2  \left. \left(\delta {Q^\dagger_L}  Q_R  \right)\right|_{0} \right\}} \, , 
\nonumber \\
\delta_{\bar{D}} ( S_{bulk} + S_{X} + S_B ) &=\displaystyle{ \int d^4x \, \left\{ \left(\int_{-\pi R^+}^{0^-} + \int_{0}^{ \pi R}\right) dy 
\ \delta {\bar{D}} \, i \Gamma^{M} \partial_{M}  D  \right.} \nonumber \\ 
& \ \ \ +\displaystyle{\left. \left. \left[- X'^*\delta D^\dagger_L Q_R + 2\,  \delta D^\dagger_R \left(D_L - \dfrac{X^*}{2} Q_L \right) \right] \right|_{\pi R} 
- 2 \left. \left(\delta {D^\dagger_R}  D_L  \right)\right|_{0} \right\}} \, .
\nonumber
\end{align}
The individual vanishing of those volume and surface terms, due to the action minimisation, 
leads to the EOM~\eqref{ELE_S1/Z2_R__Func_Free_Chiral} and the following NBC,
\begin{equation}
\left\{
\begin{array}{l}
\left. \left\{ Q_R + (X/2) \; D_R \right\} \right|_{\pi R} = 0 \, ,  \ \ \left. \left\{ D_L - (X^*/2)  \; Q_L \right\} \right|_{\pi R} = 0 \, , \\
\left. X' \; D_L \right|_{\pi R} = 0 \, ,  \ \  \left. X^{\prime *}  \; Q_R \right|_{\pi R} = 0 \, ,  \  \
\left . Q_R \right|_{0} = \left. D_L \right|_{0} = 0 \, ,
\end{array}
\right. \ 
\label{BC_S1/Z2_R__Func_Yuk_5D_BBT_1}
\end{equation}
which differ from the NBC~\eqref{BC_S1/Z2_R__Func_Yuk_NBC} obtained without the BBT. As before, given the continuity region defined by Eq.~\eqref{eq:defFP}, 
we can start by considering the search for profile solutions of 5D EOM~\eqref{ELE_S1/Z2_R__Func_Free_Chiral} and 5D NBC~\eqref{BC_S1/Z2_R__Func_Yuk_5D_BBT_1} 
on the domain, $y \in [0, \pi R]$, being equivalent to the search performed for the interval scenario (with BBT)~\cite{Angelescu:2019viv} after the replacement, 
$L \leftrightarrow \pi R$. The 4D field solutions in the decomposition~\eqref{Mix_KK_S1/Z2_R__Func} obey the known Eq.~\eqref{4D_Dirac_Mix__Func-Yuk}.
First, the opposite sign in factor of each $X^{(\prime)}$ parameter in the NBC~\eqref{BC_S1/Z2_R__Func_Yuk_5D_BBT_1}, with respect to Ref.~\cite{Angelescu:2019viv}
[see Eq.~(5.23) there], 
comes from the mentioned different Dirac matrix sign convention [{\it c.f.} $\Gamma^4$ in Eq.~\eqref{gamma_3}] and hence has no physical impact, neither on the fermion 
masses 
%[opposite signs in front of $\tan(M_n \pi R)$ in Eq.~\eqref{Normalization_1_S1/Z2_R__Func_Yuk_5D_BBT}-\eqref{Normalization_2_S1/Z2_R__Func_Yuk_5D_BBT}]
nor on the 4D effective Yukawa couplings [see Eq.~\eqref{Couplings_S1/Z2_R_Yuk_5D_BBT}]. 
Secondly, the factor $1/2$ difference at the same places in 
NBC~\eqref{BC_S1/Z2_R__Func_Yuk_5D_BBT_1}, compared to the interval NBC~\cite{Angelescu:2019viv}, comes from the existence of double numbers of surface terms 
(at $y=0$, $0^-$ and $y=\pi R$, $-\pi R^+$) -- like in Section~\ref{S1/Z2_R_Yuk_NBC__Func} -- and leads to the factor $1/2$ in the final mass 
spectrum relations~\eqref{Normalization_1_S1/Z2_R__Func_Yuk_5D_BBT}-\eqref{Normalization_2_S1/Z2_R__Func_Yuk_5D_BBT}
through a re-normalisation of the $X$ parameter as $X/2$. Thirdly, the necessary ortho-normalisation condition~\eqref{Mix_KK_Normalization_S1/Z2_R__Func_Yuk} 
can be rewritten on the domain $[0, \pi R]$ only, as [the subscript $_C$ stands for $_L$ or $_R$],
\begin{align}
\delta_{nm} \, = \, \dfrac{1}{\pi R} \int_{0}^{ \pi R} dy \, \left[ q^{n*}_{C}(y) q^m_{C}(y) + d^{n*}_{C}(y) d^m_{C}(y) \right] \, , 
\label{eq:orthonorm-on-interv} 
\end{align}
thanks to the change of variable, $y'=-y$, the fixed profile parities~\eqref{Parity_Pro_S1/Z2_R_1}-\eqref{Parity_Pro_S1/Z2_R_4} and the continuity relations~\eqref{eq:defFP}:
\begin{align}
& \int_{-\pi R^+}^{0-} dy  \left[ q^{n*}_{C}(y) q^m_{C}(y) + d^{n*}_{C}(y) d^m_{C}(y) \right] 
= \int_{0^+}^{\pi R^-} dy'  \left[ q^{n*}_{C}(-y') q^m_{C}(-y') + d^{n*}_{C}(-y') d^m_{C}(-y') \right] 
\nonumber \\ & \ \ \ \ \ \ 
=\int_{0^+}^{\pi R^-} dy'  \left[ q^{n*}_{C}(y') q^m_{C}(y') + d^{n*}_{C}(y') d^m_{C}(y') \right] 
= \int_{0}^{\pi R} dy  \left[ q^{n*}_{C}(y) q^m_{C}(y) + d^{n*}_{C}(y) d^m_{C}(y) \right] \, , \nonumber 
\end{align}
recovering thus exactly and conveniently the interval condition, if $L = \pi R$. Nevertheless, including the factor $1/\sqrt{2\pi R}$, 
the dimensional wave functions [mass dimension $1/2$] are identical within the orbifold and interval frameworks
only up to an additional normalisation factor $1/\sqrt{2}$ here, due to the double compact space size: see the respectively used decomposition 
normalisations~\eqref{Mix_KK_S1/Z2_R__Func} above and (4.1) in Ref.~\cite{Angelescu:2019viv}.
Therefore, we can finally apply the results of Ref.~\cite{Angelescu:2019viv} here for the SM-like consistent solutions of the fields over $y \in [0, \pi R]$:
we find, for the dimensionless profiles ($\forall \, n \in \mathbb{N}$), 
\begin{equation}
\left\{
\begin{array}{l}
(+\times): \ q_{L}^n(y) = A^n_q \, \cos(M_n \, y) \, ,  \ \ \ (-\times): \ q_{R}^n(y) = - A^n_q \, \sin(M_n \, y) \, , \\
(-\times): \ d_{L}^n(y) =  A^n_d \, \sin(M_n \, y) \, ,  \ \ \ (+\times): \ d_{R}^n(y) = A^n_d \, \cos(M_n \, y) \, ,
\end{array}
\right. \ 
\label{Profiles_S1/Z2_R__Func_Yuk_EBC}
\end{equation}
for the two classes of real mass spectrum solutions ($X = |X| \, e^{i\alpha_Y}$ with $\alpha_Y, \alpha_0^n \in \mathbb{R}$),
\begin{eqnarray} 
&  \ \tan(M_n \; \pi R) =  \, \left|\dfrac{X}{2} \right|, \ A_q^n = e^{i(\alpha_0^n+\alpha_Y)}, 
\ A_d^n = e^{i\alpha_0^n}, 
\label{Normalization_1_S1/Z2_R__Func_Yuk_5D_BBT}
\\ 
& \ \tan(M_n \; \pi R) = - \, \left|\dfrac{X}{2} \right|, \ A_q^n = e^{i(\alpha_0^n+\alpha_Y\pm\pi)}, 
\ A_d^n = e^{i\alpha_0^n}, 
\label{Normalization_2_S1/Z2_R__Func_Yuk_5D_BBT}
\end{eqnarray}
and for the absolute values of the fermion masses [based on Eq.~\eqref{ntilde}],
\begin{equation}
|M_n| = \dfrac{1}{\pi R} \left \vert \, \arctan \left \vert \dfrac{X}{2} \, \right \vert + (-1)^n \, \tilde n(n) \,  \pi \, \right \vert  \, .
\label{MS_S1/Z2_R__Func_Yuk_5D_BBT_2}
\end{equation}
We call $(\times)$ the new Yukawa coupling (in $X$) dependent BC, given by Eq.~\eqref{Profiles_S1/Z2_R__Func_Yuk_EBC}, 
\eqref{Normalization_1_S1/Z2_R__Func_Yuk_5D_BBT}-\eqref{Normalization_2_S1/Z2_R__Func_Yuk_5D_BBT}, \eqref{MS_S1/Z2_R__Func_Yuk_5D_BBT_2} 
at the brane located at $y=\pi R$, in order to distinguish them from the Dirichlet BC usually noted $(-)$ and the Neumann BC $(+)$.
Note that, similarly to the free solutions~\eqref{completeEBCsm0}, the opposite signs in front of the $(-\times)$ 
profiles~\eqref{Profiles_S1/Z2_R__Func_Yuk_EBC}, with respect to the results of Ref.~\cite{Angelescu:2019viv},
simply come from a different sign convention for the $\Gamma^4$ matrix.
At this stage, the part of the profile solutions on the complementary region, $y\in [-\pi R^+, 0^-]$, is deduced through the four types of 
$\mathbb{Z}_2$ transformations~\eqref{Parity_Pro_S1/Z2_R_1}-\eqref{Parity_Pro_S1/Z2_R_4}. Hence, the $M_n$ spectrum entering the profile 
solutions in both regions, $[0, \pi R]$ and $[-\pi R^+, 0^-]$, is the same. As a first conclusion, the introduction of the BBT allows to obtain 
realistic fermion wave functions and consistent mass eigenvalues. The absolute mass spectrum obtained within the 5D approach in  
Eq.~\eqref{MS_S1/Z2_R__Func_Yuk_5D_BBT_2}  is analytically matching the one derived via the 4D method in Eq.~\eqref{MS_S1/Z2_R__Func_Yuk_4D} 
for a real Yukawa coupling constant: this feature represents a non-trivial confirmation of the present exact results. In particular, the absence of $X'$ parameter in the fermion 
4D mass matrix $\mathcal{M}$, described below Eq.~\eqref{Coeff_S1/Z2_R_4D__Func}, is interestingly recovered through the mass independence from $X'$ as induced 
as well by the condition $$X'=0 \, ,$$ issued from the 5D NBC~\eqref{BC_S1/Z2_R__Func_Yuk_5D_BBT_1}. Regarding the probability current, the component $j^4\vert_{\pi R}$ 
at the Yukawa brane is still given by Eq.~\eqref{Current_BC_S1/Z2_R__Func_Yuk_EBC-Yuk} since the BBT do not affect it, as explained at the end of 
Section~\ref{S1/Z2_R_Free_BBT__Func}. The relations found in the first line of the NBC~\eqref{BC_S1/Z2_R__Func_Yuk_5D_BBT_1}, injected once into each
term of this current component expression, give rise to, $$j^4\vert_{\pi R} \, = \, 0 \,  .$$
The BBT are thus found to induce NBC leading to a vanishing current component along the extra dimension at the fixed points of the orbifold, 
with (present section) or without ({\it c.f.} Section~\ref{S1/Z2_R_Free_BBT__Func}) a brane-localised Yukawa coupling, and in turn to a continuous 
current component along the extra dimension at those points given the odd parities, demonstrated in Eq.~\eqref{Current_Parity_S1/Z2_R__Func_Free_EBC-Yuk} 
or \eqref{Current_Parity_S1/Z2_R__Func_Free_EBC} respectively.

\vspace{0.9cm}
\begin{table}[h]
\centering
\resizebox{\textwidth}{!}
{
\begin{tabular}{| c | c | c | c | c | c | }
\hline
 & \multirow{4}{*}{$\mathbb{Z}_2$} & \multicolumn{4}{ c |}{Fields} \\
\cline{3-6}
{Continuity} &  & \multicolumn{2}{ c |}{$Q_{L/R}$} & \multicolumn{2}{ c |}{$D_{L/R}$}  \\
\cline{3-6}
{domains} 
 &  & \multirow{2}{*}{$q^n_L(y)/(\pm e^{i(\alpha_0^n+\alpha_Y)})$} & \multirow{2}{*}{$q^n_R(y)/(\pm e^{i(\alpha_0^n+\alpha_Y)})$} 
 & \multirow{2}{*}{$d^n_L(y)/e^{i\alpha_0^n}$} & \multirow{2}{*}{$d^n_R(y)/e^{i\alpha_0^n}$} \\
& & & & & \\
\hline
$[0, \pi R]$ & Any & $  \cos(M_n \; y)$ & $-\sin(M_n \; y)$ & $  \sin(M_n \; y)$ & $   \cos(M_n \; y)$ \\
\hline
\multirow{4}{*}{$[-\pi R^+, 0^-]$} & \uppercase\expandafter{\romannumeral1} & $   \cos(M_n \; y)$ & $ -\sin(M_n \; y)$ & $  \sin(M_n \; y)$ & $   \cos(M_n \; y)$ \\
\cline{2-6}
& \uppercase\expandafter{\romannumeral2} & $-    \cos(M_n \; y)$ & $   \sin(M_n \; y)$ & $ - \sin(M_n \; y)$ & $-   \cos(M_n \; y)$ \\
\cline{2-6}
& \uppercase\expandafter{\romannumeral3} & $   \cos(M_n \; y)$ & $ - \sin(M_n \; y)$ & $ -\sin(M_n \; y)$ & $-   \cos(M_n \; y)$ \\
\cline{2-6}
& \uppercase\expandafter{\romannumeral4} & $-    \cos(M_n \; y)$ & $   \sin(M_n \; y)$ & $   \sin(M_n \; y)$ & $   \cos(M_n \; y)$ \\
\hline\hline
KK Masses & \multicolumn{5}{ c |}{$|M_n| =  \left \vert \arctan \left \vert X/2 \, \right \vert + (-1)^n \, \tilde n(n) \, \pi \right \vert / \pi R$, $n \in \mathbb{N}$}  \\
\hline
\end{tabular}
}
\vspace{0.25cm}
\caption{SM-like coupled fermion profiles on the two orbifold continuity 
domains $[-\pi R^+, 0^-]$ and $[0, \pi R]$, corresponding to the solutions~\eqref{Profiles_S1/Z2_R__Func_Yuk_EBC}, 
\eqref{Normalization_1_S1/Z2_R__Func_Yuk_5D_BBT}-\eqref{Normalization_2_S1/Z2_R__Func_Yuk_5D_BBT}, together with the associated
absolute mass spectrum~\eqref{MS_S1/Z2_R__Func_Yuk_5D_BBT_2} for completeness. The profiles are given for the four types of $\mathbb{Z}_2$ 
transformations~\eqref{Parity_Pro_S1/Z2_R_1}-\eqref{Parity_Pro_S1/Z2_R_4}.}
\label{tab:SMQ&D_Coupled_Z2}
\end{table}
\vspace{0.4cm}

In Table~\ref{tab:SMQ&D_Coupled_Z2} are exhibited the explicit profile functions over the entire orbifold domain for the SM-like 
solutions~\eqref{Profiles_S1/Z2_R__Func_Yuk_EBC}, \eqref{Normalization_1_S1/Z2_R__Func_Yuk_5D_BBT}-\eqref{Normalization_2_S1/Z2_R__Func_Yuk_5D_BBT},
\eqref{MS_S1/Z2_R__Func_Yuk_5D_BBT_2}. We can see on this table that the choice of type of $\mathbb{Z}_2$ transformation is purely a convention because
it can modify the profile signs but without effects on the mass spectrum.

\begin{figure}[h]
\centering
\includegraphics[width=15cm]{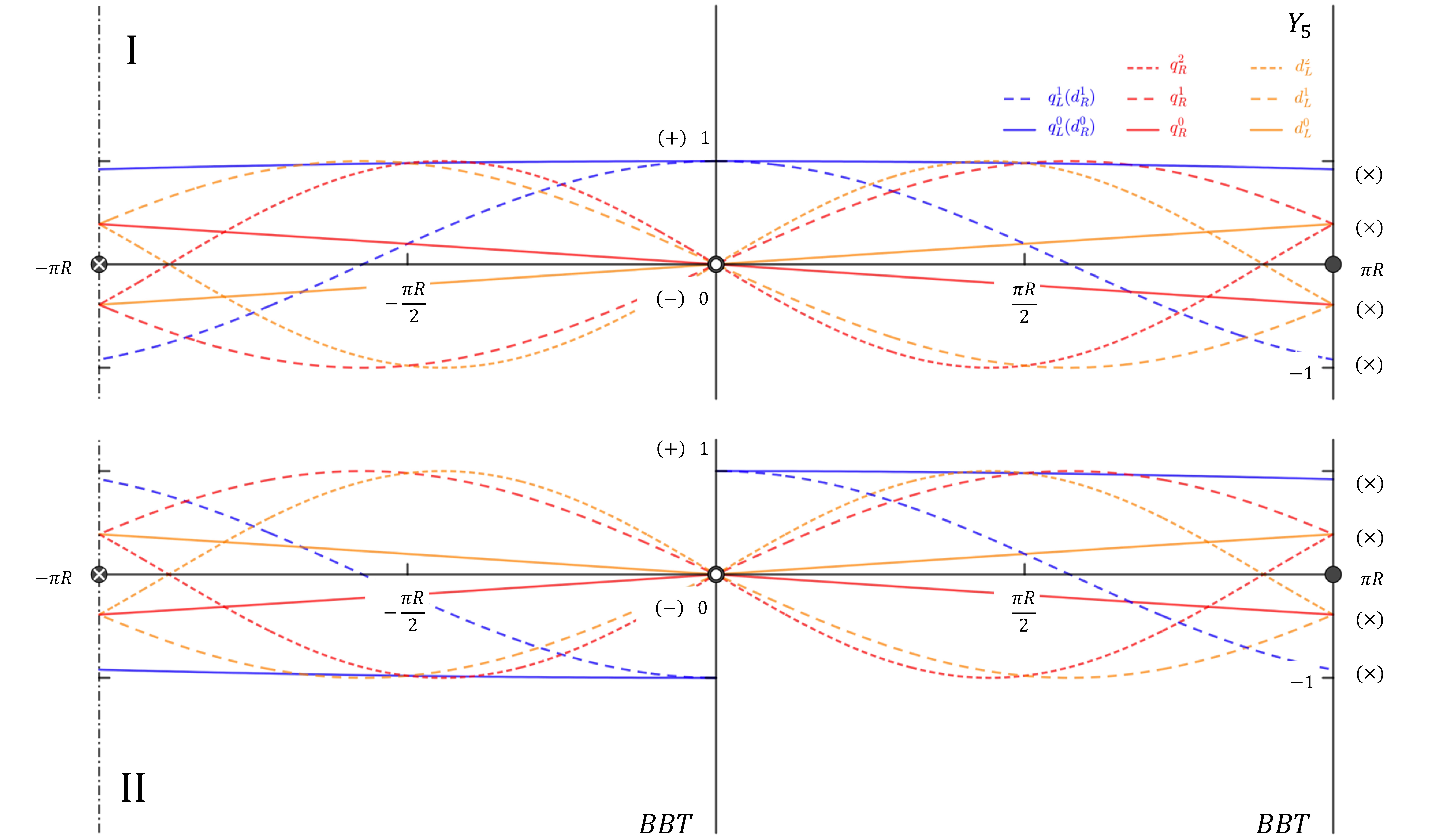}
\vspace{0.25cm}
\caption{Zero-mode and excitation wave functions $q^n_{L/R}(y)$, $d^n_{L/R}(y)$, with $n=0,1,2$, along the $\mathcal{S}^1/ \mathbb{Z}_2$ orbifold domain, 
$y \in [-\pi R^+, 0^-] \cup [0, \pi R]$, corresponding to the Yukawa-coupled solutions~\eqref{Normalization_1_S1/Z2_R__Func_Yuk_5D_BBT}, presented in
Table~\ref{tab:SMQ&D_Coupled_Z2}, for the simplified case, $\alpha_Y=\alpha^n_{0}=0$, and the two different types of $\mathbb{Z}_2$ transformations, 
$\uppercase\expandafter{\romannumeral1}$, $\uppercase\expandafter{\romannumeral2}$ from Eq.~\eqref{Parity_Pro_S1/Z2_R_1}-\eqref{Parity_Pro_S1/Z2_R_4}. 
The two fixed points at, $y=0$, $y=\pi R\equiv -\pi R$, the BC, $(-)/(+)/(\times)$, the BBT and Yukawa coupling brane-locations are indicated on the graph.}
\label{Profiles_S1/Z2_R_Yuk_Graphic}
\end{figure}

In Figure~\ref{Profiles_S1/Z2_R_Yuk_Graphic}, we illustrate a set of excitation profiles, obeying the $\mathbb{Z}_2$ transformations of types 
$\uppercase\expandafter{\romannumeral1}$ and $\uppercase\expandafter{\romannumeral2}$ in Eq.~\eqref{Parity_Pro_S1/Z2_R_1}-\eqref{Parity_Pro_S1/Z2_R_4}, 
for the found Yukawa-coupled solutions~\eqref{Normalization_1_S1/Z2_R__Func_Yuk_5D_BBT}, 
which are explicitly presented in Table~\ref{tab:SMQ&D_Coupled_Z2}, within the simplified real case, $\alpha_Y=\alpha^n_{0}=0$. 
We observe on this figure that all the wave function values at the Yukawa-brane (at the fixed point, $y=\pi R$) are modified due to the presence of this coupling.
For example, under the type~$\uppercase\expandafter{\romannumeral1}$ of $\mathbb{Z}_2$ transformation, the profile values $d^n_L(\pi R)=d^n_L(\pi R^-)$  
are shifted from zero as well as from $d^n_L(-\pi R^+)$, in contrast with the free case shown in Fig.~\ref{Profiles_S1/Z2_R_Free_Graphic}. This shift creates profile jumps 
whose amplitude is depending on the Yukawa coupling constant through the $X$ parameter [BC $(\times)$ from Eq.~\eqref{Profiles_S1/Z2_R__Func_Yuk_EBC}, 
\eqref{Normalization_1_S1/Z2_R__Func_Yuk_5D_BBT}-\eqref{Normalization_2_S1/Z2_R__Func_Yuk_5D_BBT}, \eqref{MS_S1/Z2_R__Func_Yuk_5D_BBT_2}].
Under the type~$\uppercase\expandafter{\romannumeral2}$ of $\mathbb{Z}_2$ transformation, the same figure shows clearly that the profile 
jump $d^n_L(\pi R)=d^n_L(\pi R^-)\neq d^n_L(-\pi R^+)$ disappears but then other kinds of jump arise like: $q^n_L(\pi R)=q^n_L(\pi R^-)\neq q^n_L(-\pi R^+)$ and
$q^n_L(0^-) \neq q^n_L(0) = q^n_L(0^+)$.
The presence of new possible profile discontinuities justifies once more mathe\-matically the prescriptions about the field continuities 
and action integration domains introduced in Section~\ref{5D_Geo}.

\vspace{0.5cm}

Finally, let us calculate, still without any kind of Higgs field regularisation, 
the physical 4D effective Yukawa coupling constants between the mass eigenstates $\psi_L^n(x^\mu)$ and $\psi_R^m(x^\mu)$ as generated by the insertion of 
decompositions~\eqref{Mix_KK_S1/Z2_R__Func} into Eq.~\eqref{S_h_S1/Z2_R__Func}, based on the obtained profile expressions~\eqref{Profiles_S1/Z2_R__Func_Yuk_EBC},
\eqref{Normalization_1_S1/Z2_R__Func_Yuk_5D_BBT}-\eqref{Normalization_2_S1/Z2_R__Func_Yuk_5D_BBT}, \eqref{MS_S1/Z2_R__Func_Yuk_5D_BBT_2}:
\begin{align}
y_{nm} &  \, \hat = -\dfrac{Y_5}{2\sqrt{2} \pi R} \, q_L^{n*} (\pi R) \; d_R^m(\pi R) \nonumber \\
 &= \mp \dfrac{|Y_5|}{2\sqrt{2} \pi R} \, e^{i(\alpha_0^m-\alpha_0^n)} \cos(M_n \; \pi R) \cos(M_m \; \pi R) \nonumber \\
 &= \mp (-1)^{\tilde n(n)+\tilde n(m)} \, e^{i(\alpha_0^m-\alpha_0^n)} \, \dfrac{|Y_5|}{2\sqrt{2} \pi R(1+|X/2|^2)} \, ,  
\label{Couplings_S1/Z2_R_Yuk_5D_BBT}
\end{align}
where we have used a trigonometric identity~\footnote{For $n \in \mathbb{Z}$, one has, $\cos(\theta + n \pi) 
= (-1)^n \cos(\theta)$, and for $T \in \mathbb{R}$, $\cos[\arctan(T)] = \dfrac{1}{\sqrt{1+T^2}}$.} 
to get the last equality. In the decoupling limit of extremely heavy KK modes, $R\rightarrow 0$, we can then
write the modulus of the lightest mode coupling constant, using Eq.~\eqref{eq:RefDecouplY5}, as,
\begin{equation}
\vert y_{00} \vert \mathop{\to} \limits_{\substack{R\rightarrow 0}} \, \frac{|Y_5|}{2\sqrt{2} \pi R} = \frac{|y_4|}{\sqrt{2}}\, , 
\ {\rm since,} \ X=\frac{v}{\sqrt{2}} \, Y_5 = \frac{v }{\sqrt{2}} \, 2\pi R \, y_4 \, ,
\label{MS_DL_Interval & L_N__Func_Yuk_5D}
\end{equation}
and the absolute mass eigenvalue of the lightest eigenstates as [from Eq.~\eqref{MS_S1/Z2_R__Func_Yuk_5D_BBT_2}],
\begin{equation}
\vert M_0 \vert \mathop{\to} \limits_{\substack{R\rightarrow0}} \, \dfrac{\vert X \vert}{2\pi R} = \dfrac{v \vert Y_5 \vert}{2\sqrt{2} \pi R} 
\, \mathop{\to} \limits_{\substack{R\rightarrow0}} \,  v \, \vert y_{00} \vert \, ,
\label{MS_DL_S1/Z2_R__Func_Yuk_5D_BBT}
\end{equation}
so that the SM fermion set-up -- for the assumed single family -- is recovered as expected from the decoupling condition. Besides, 
we can conclude that the choice of type of $\mathbb{Z}_2$ transformation among Eq.~\eqref{Parity_Pro_S1/Z2_R_1}-\eqref{Parity_Pro_S1/Z2_R_4} affects 
neither the profile values taken at the point $y=\pi R$ -- see Table~\ref{tab:SMQ&D_Coupled_Z2} --
nor their global ortho-normalisation condition~\eqref{Mix_KK_Normalization_S1/Z2_R__Func_Yuk} --
as described right below Eq.~\eqref{eq:orthonorm-on-interv} -- so that the 4D effective Yukawa coupling constants~\eqref{Couplings_S1/Z2_R_Yuk_5D_BBT} 
are insensitive as well to this $\mathbb{Z}_2$ representation choice.

%((\section{From fields as distributions to fields as functions}
%\label{fields_distrib}
%\input{Field_distrib.tex}))

\section{The inclusive $\mathbb{Z}_2$ parity}
\label{subsec:Z2inclusive}

Let us study the alternative scenario whose definition is based on the $\mathbb{Z}_2$ transformation of 5D fields extended to include the two fixed points
at $y=0$ and $y=\pi R$:
\begin{equation}
\Phi(x^\mu, -y)={\mathcal T} \Phi(x^\mu, y) \, , \ \ \forall y \in (-\pi R, \pi R] \, ,
\label{eq:5DFtrans-incl}
\end{equation}
in contrast with Eq.~\eqref{eq:5DFtrans}. This generic transformation still lets the Lagrangian density invariant, exactly like in Eq.~\eqref{eq:Z2Lag}.
At the two fixed points, this Lagrangian invariance is once more automatically satisfied without the need for any specific ${\mathcal T}$ transformation.
Accordingly to the simple Eq.~\eqref{eq:5DFtrans-incl}, the operator ${\mathcal T}$ for the fixed points is the same as the non-trivial one which must let the Lagrangian 
invariant in the bulk. Let us consider in particular the realistic $\mathbb{Z}_2$ transformation leading to the SM chirality set-up: it is the bulk transformation in 
Eq.~\eqref{Parity_S1/Z2_R_1}, defined now over the same range as in Eq.~\eqref{eq:5DFtrans-incl}, which keeps well $\mathcal{L}_{kin}$ invariant in the bulk 
according to Eq.~\eqref{eq:Z2Lag}:
\begin{equation} 
\left\{
\begin{array}{c c c}
Q \left( x^\mu, -y \right) = - \gamma^5 \, Q \left( x^\mu, y \right)  
\\ \vspace{-0.2cm} \\
D \left( x^\mu, -y \right) = \gamma^5 \, D \left( x^\mu, y \right) 
\end{array}
\right. , \  \  \  \forall y \in (-\pi R, \pi R] \, .
\label{Parity_S1/Z2_R_1-incl}
\end{equation}
Focusing on the fixed points at $y=0$ and $y=\pi R \equiv - \pi R$, we obtain the four non-trivial relations 
\begin{eqnarray} 
\left\{
\begin{array}{c c c}
Q \left( x^\mu, 0 \right) = - \gamma^5 \, Q \left( x^\mu, 0 \right)   \, \Rightarrow \, Q_L \vert_0 = Q_L  \vert_0 \, , \, {\rm and,} \ Q_R  \vert_0 = - Q_R  \vert_0 = 0
\\ \vspace{-0.2cm} \\
D \left( x^\mu, 0 \right) = \gamma^5 \, D \left( x^\mu, 0 \right)  \, \Rightarrow \, D_L  \vert_0 = - D_L  \vert_0 = 0 \, , \, {\rm and,} \ D_R \vert_0 = D_R \vert_0
\end{array}
\right.
\label{EBCprime} \\  \nonumber\\
\left\{
\begin{array}{c c c}
Q \left( x^\mu, \pi R \right) = - \gamma^5 \, Q \left( x^\mu, \pi R \right)   \, \Rightarrow \, Q_R  \vert_{\pi R} = 0
\\ \vspace{-0.2cm} \\
D \left( x^\mu, \pi R \right) = \gamma^5 \, D \left( x^\mu, \pi R \right)  \, \Rightarrow \, D_L  \vert_{\pi R} = 0  
\end{array}
\right.  \  \  \   \  \  \   \  \  \   \  \  \  \  \  \   \  \  \   \  \  \    {\rm [EBC']}
\nonumber 
\end{eqnarray}
representing new EBC that we denote EBC' to distinguish them from those in Eq.~\eqref{Gen_BC_S1/Z2_R__Func_Free_EBC}.

In the free case, Section~\ref{S1/Z2_R_Free_NBC__Func} has shown that EBC(') or BBT must be considered.
Starting with the EBC('), in analogy with Section~\ref{S1/Z2_R_Free_EBC__Func}, the fixed $\mathbb{Z}_2$ transformations~\eqref{Parity_S1/Z2_R_1-incl} 
in the bulk lead to the EBC~\eqref{Gen_BC_S1/Z2_R__Func_Free_EBC} while the $\mathbb{Z}_2$ transformations~\eqref{EBCprime} at the fixed points lead to the EBC'. 
Those EBC' select one general BC set among these four EBC sets for the 5D field $Q$, and same statement for $D$: the sets corresponding to the chiral
solution of line~1 (2) in Eq.~\eqref{completeEBCsm0} for the field $D$ ($Q$), namely the SM-like chirality configuration. Finally, the 
complete profile solutions over the whole orbifold domain are found out as before via the bulk $\mathbb{Z}_2$ transformations~\eqref{Parity_S1/Z2_R_1-incl}.

Alternatively, the selected consistent BBT~\eqref{S_BBT_S1/Z2_R__Func} can be included like in Section~\ref{S1/Z2_R_Free_BBT__Func} to obtain the same SM-like solutions. 
The corresponding EBC'~\eqref{EBCprime}, part of the EBC~\eqref{Gen_BC_S1/Z2_R__Func_Free_EBC} and required by the model, are checked to be satisfied
afterwards, as consequences.

Once the free profiles are worked out as described right above -- either through the EBC(') or the BBT -- 
we can apply the 4D method of Section~\ref{S1/Z2_R_Yuk_4D__Func}, based on infinite matrix diagonalisation, 
in order to derive the mass spectrum in the presence of brane-localised Yukawa couplings.
Even the 4D effective Yukawa coupling constants can be calculated in this way: the above EBC' selection of a specific chirality set-up and mass
spectrum for the free fields would affect as well these effective coupling constants, for instance via the KK mass mixings.

In contrast, the analysis of point-like Yukawa interactions cannot be achieved via the 5D approach within the present inclusive $\mathbb{Z}_2$ symmetry model.

First, the EBC(') motivated by Section~\ref{S1/Z2_R_Yuk_NBC__Func} must be split into the EBC coming directly from the vanishing probability currents -- or 
say indirectly from 
the fixed $\mathbb{Z}_2$ transformations~\eqref{Parity_S1/Z2_R_1-incl} in the bulk -- discussed in Section~\ref{S1/Z2_R_Yuk_EBC2__Func}
and the EBC'~\eqref{EBCprime}. These EBC' combined with the surface terms at $y=\pi R$ in 
Eq.~\eqref{S_D_S1/Z2_R__Func_Yuk_HVP_EBC-Yuk}, including the Yukawa terms, give rise to the BC
of type~\eqref{BC_S1/Z2_R__Func_Yuk_NBC} involving only single terms proportional to the Yukawa coupling constant and equal to zero. Hence, the 
resulting mass spectrum looses its dependence on the Yukawa coupling constant which conflicts with the decoupling limit argument
[see Eq.~\eqref{MS_DL_S1/Z2_R__Func_Yuk_5D_BBT}]. 

Secondly, the BBT~\eqref{S_BBT_S1/Z2_R__Func} could be added like in Section~\ref{S1/Z2_R_Yuk_BBT__Func} to try obtaining SM-like solutions. 
However the EBC'~\eqref{EBCprime}, expected to be recovered afterwards, are not compatible with the resulting BC~\eqref{Profiles_S1/Z2_R__Func_Yuk_EBC}
together with the spectrum equations~\eqref{Normalization_1_S1/Z2_R__Func_Yuk_5D_BBT}-\eqref{Normalization_2_S1/Z2_R__Func_Yuk_5D_BBT}.

\section{Result analysis}
\label{S1/Z2_R_LM__Func}

\subsection{The higher-dimensional method}
 
\begin{figure}[!h]
\begin{center}
\vspace{1cm}
\includegraphics[width=10cm]{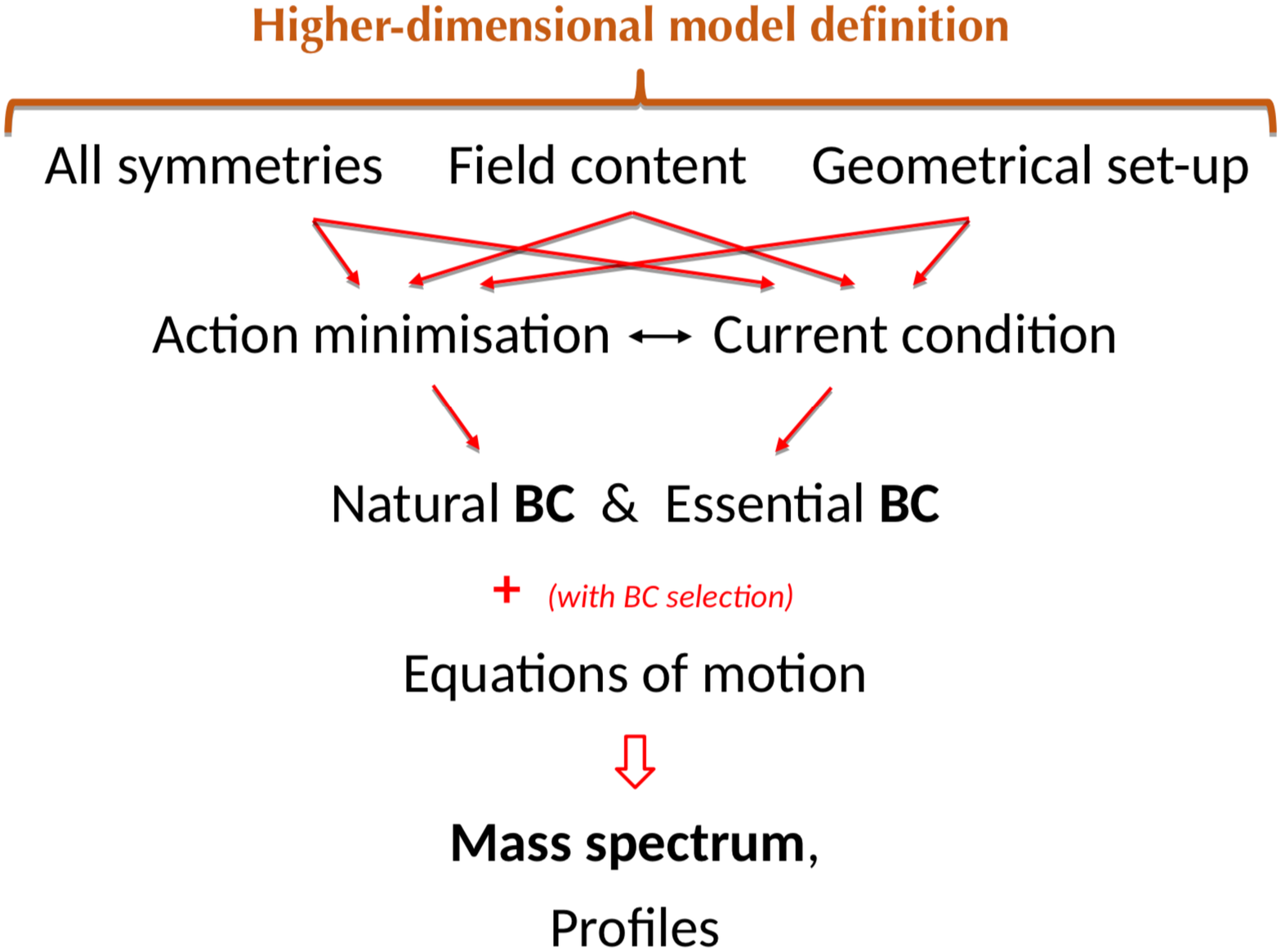}
%\vspace{-0.5cm}
\end{center}
\caption{Schematic inverse pyramidal picture describing the generic process for determining the mixed KK mass spectrum and bulk field wave functions.
Acronym notations are the same as in the main text.}
\label{fig:Pyramidal}
\end{figure}

The present study confirms the general methodology depicted in Fig.~\ref{fig:Pyramidal} and presented in Ref.~\cite{Angelescu:2019viv}. 
Within the present model, the probability current condition on this schematic description is the vanishing of fermion currents 
at the two fixed points (issued from $\mathbb{Z}_2$ symmetry criteria and inducing the EBC~\eqref{Gen_BC_S1/Z2_R__Func_Free_EBC} in the free case).
For the interval model, the vanishing current condition is a direct implication of the existence of boundaries for the matter fields.
This current vanishing holds both in the presence and absence of brane-localised Yukawa couplings.

In the framework of the orbifold version described in Section~\ref{subsec:Z2inclusive}, the additional field condition~\eqref{EBCprime}, coming from the $\mathbb{Z}_2$ symmetry at the fixed points, 
accompanies the definition of the $\mathbb{Z}_2$ symmetry of the bulk action and leads to the new EBC'.

\subsection{Discussion of the action content}

In addition to the information contained in the action~\eqref{1_eq:actionTot}, the present orbifold model is defined in a complementary way by other elements like: 
{\it (i)} the $\mathcal{S}^1$ junction point at $y = \pi R \equiv -\pi R$,
{\it (ii)} the choices of $\mathbb{Z}_2$ transformations for the fields in the bulk [see Eq.~\eqref{Parity_S1/Z2_R_1}-\eqref{Parity_S1/Z2_R_4}] and possibly at the fixed points [{\it c.f.} Eq.~\eqref{EBCprime}],
{\it (iii)} the EBC~\eqref{Gen_BC_S1/Z2_R__Func_Free_EBC} imposed by the model definition when those are used instead of the BBT.
Regarding the point {\it (iii)}, Table~\ref{ST_S1/Z2_R__Func} summarises the obtained cases where the EBC and the BBT can be used.
This table is identical to the one obtained in the interval model study~\cite{Angelescu:2019viv}.

\vspace{0.5cm}
\begin{table}[h]
\centering
\begin{tabular}{ c  c  c  c }
\multicolumn{1}{m{2.7cm}<\centering |}{}& \multicolumn{1}{m{3cm}<\centering |}{\bf No boundary characteristic} & 
\multicolumn{1}{m{3.7cm}<\centering |}{\bf Vanishing current condition [EBC]} & \multicolumn{1}{m{3cm}<\centering}{\bf Bilinear brane terms [NBC]} \\
\hline
\multicolumn{1}{m{3cm}<\centering |}{\bf 4D Approach} & \multicolumn{1}{m{3cm}<\centering |}{\it (Impossible)} & \multicolumn{1}{m{3cm}<\centering |}{BC ($\pm$)} & \multicolumn{1}{m{3cm}<\centering}{BC ($\pm$)} \\
\hline
\multicolumn{1}{m{3cm}<\centering |}{\bf 5D Approach} & \multicolumn{1}{m{3cm}<\centering |}{\it (Impossible)} & \multicolumn{1}{m{3cm}<\centering |}{\it (Impossible)} & \multicolumn{1}{m{3cm}<\centering}{BC ($\times$)} 
\end{tabular}
\vspace{0.25cm}
\caption{Types of boundary conditions for the bulk fermions at an orbifold fixed point where is located their interactions with the Higgs boson, in different brane treatments:
presence of BBT, vanishing of probability current or nothing specific. The 4D line holds as well for the 5D approach of the free brane. As usually, the Dirichlet BC are noted $(-)$, 
the Neumann BC $(+)$ and we denote $(\times)$ the new BC depending on the Yukawa coupling constant [corresponding to Eq.~\eqref{Profiles_S1/Z2_R__Func_Yuk_EBC} taken at $y=\pi R$]. 
The (N,E)BC acronym definitions are the same as in the text.}
\label{ST_S1/Z2_R__Func}
\end{table}

\subsection{About the orbifold/interval duality}

The present $\mathcal{S}^1/\mathbb{Z}_2$ orbifold model and the $[0,L]$ interval scenario studied in Ref.~\cite{Angelescu:2019viv} are physically different:
their geometrical set-ups and Lagrangian symmetries are not identical. Nevertheless, the respective theoretical predictions for the observables like 
the (brane-coupled) fermion mass spectra and 4D effective Yukawa coupling constants are identical up to factors $2$, which may be called a duality. 
Indeed, for a comparable dimension size $L=\pi R$, although the mass absolute values~\eqref{MS_S1/Z2_R__Func_Yuk_5D_BBT_2} involve 
a new factor $1/2$ in front of $X$, with respect to the interval analytical result, the measurable range of values for $\vert M_n \vert $ is of the same order and 
the precise limits of this range rely on the approximate perturbative limits of the 4D effective Yukawa coupling constants proportional to $y_4$
[see Sections~\ref{1_Yukawa} and 
Eq.~\eqref{Couplings_S1/Z2_R_Yuk_5D_BBT}-\eqref{MS_DL_Interval & L_N__Func_Yuk_5D}-\eqref{MS_DL_S1/Z2_R__Func_Yuk_5D_BBT}]. 
Besides, the dependence of the analytical mass formula on the Lagrangian parameters
is identical in the two models, up to this factor $1/2$ entering the coupling constant definition, 
as can be seen from Eq.~\eqref{MS_S1/Z2_R__Func_Yuk_5D_BBT_2} and Section~\ref{1_Yukawa} -- including the free limiting case $X\to 0$.
Similar comments hold for the 4D effective Yukawa coupling constants~\eqref{Couplings_S1/Z2_R_Yuk_5D_BBT} which have additional factors $1/2$
in front of $X$ and as an overall factor (latter one induced by normalisation considerations), with respect to the interval case.

The orbifold version of Section~\ref{subsec:Z2inclusive} contains additional information at the fixed point branes. It predicts thus 
a specific chirality configuration and mass spectrum [among chiral or vector-like solutions respectively of type~\eqref{completeEBCsm0}-\eqref{completeEBCcusto0}] 
so that it is not dual to the interval model.

Coming back to the case of duality, there exist similarities between the orbifold and interval models, 
as it appeared throughout this work when solving the EOM and BC to find out the fields. Let us now comment on the similarities at the Lagrangian level.
First, the BBT~\eqref{S_BBT_S1/Z2_R__Func} have the same form as in the interval framework~\cite{Angelescu:2019viv} and the different factor $2$ 
is related to the double size of the compactified space for the identification, $L=\pi R$. The opposite front sign in the BBT (for a similar profile solution set-up) 
is just due to a different Dirac matrix sign convention [see $\Gamma^4$ sign in Eq.~\eqref{gamma_3}].

In the global action~\eqref{1_eq:actionTot}, $S_{bulk}$ remains to be discussed, the other parts being identical in the orbifold and interval models.
Thanks to the orbifold property~\eqref{eq:Z2Lag}, the change of variable, $y'=-y$, allows the following rewriting of Eq.~\eqref{eq:Sstart},
\begin{eqnarray}
S_{bulk} & = &\int d^4x \left ( \int_{-\pi R^+}^{0^-} dy\ \mathcal{L}_{kin}(y) + \int_0^{\pi R} dy\ \mathcal{L}_{kin}(y) \right )  
\label{eq:Sstart-dual}  \\ & = &\int d^4x \left ( \int^{\pi R^-}_{0^+} dy'\ \mathcal{L}_{kin}(y') + \int_0^{\pi R} dy\ \mathcal{L}_{kin}(y) \right ) 
= 2 \, \int d^4x \left ( \int_0^{\pi R} dy\ \mathcal{L}_{kin}(y) \right )  ,
\nonumber
\end{eqnarray}
where the last step is based on Eq.~\eqref{eq:defFP}. Therefore, using Eq.~\eqref{1_eq:actionTot} and the relevant identification, $L=\pi R$, 
we can express the orbifold action in terms of the interval action pieces~\cite{Angelescu:2019viv} (indicated by the $L$ exponent):
\begin{eqnarray}
S_{5D} & = & 2 \, S^L_{bulk} + S^{(L)}_H + S^{(L)}_X + S^{(L)}_{int}  + 2 \, S^L_B 
\nonumber \\ & = & 2 \, \left [ S^L_{bulk} + \frac{1}{2} \{ S^{(L)}_X + S^{(L)}_{int} \} + S^L_B \right ] + S^{(L)}_H 
\label{1_eq:actionTot-interval} \\ & = & 2 \, \left [ S^L_{bulk} +  S^{(L)}_{X/2} + S^{(L)}_{int}\vert_{X/2} + S^L_B \right ] + S^{(L)}_H  \, . 
\nonumber 
\end{eqnarray}
This re-expression reveals an alternative method to derive the fermion masses and couplings, which are independent from the pure scalar part, namely $S^{(L)}_H$.
The idea is that, within the orbifold model now described by the action~\eqref{1_eq:actionTot-interval} importantly together with the description of the $\mathbb{Z}_2$ symmetry 
over $\mathcal{S}^1$, we can first search for the field parts along the limited domain $[0,\pi R]$. This search is in fact based on the action 
$[ S^L_{bulk} $ + $  S^{(L)}_{X/2} $ + $ S^{(L)}_{int}\vert_{X/2} $ + $ S^L_B ]$, since the overall factor $2$ in Eq.~\eqref{1_eq:actionTot-interval} 
affects neither the EOM (global factor) nor the BC (same factor in front of the surface terms and pure brane terms combined into BC)~\footnote{This search could also 
be constrained by vanishing currents at $y=0,\pi R$ instead of the $S^L_B$ presence, in the free case, as shown in Sections~\ref{S1/Z2_R_Free_EBC__Func} and 
\ref{S1/Z2_R_Yuk_EBC2__Func}.}, and is in turn strictly equivalent to solving the interval model. Given this action, the solutions obtained for the 4D masses
(and 4D effective Yukawa coupling constants from profile overlaps with the Higgs boson peak at $y=\pi R$) 
are those of Ref.~\cite{Angelescu:2019viv} but involving a normalised coupling parameter $X/2$. The last stage of this technics is the extension 
of the obtained profiles over the complete orbifold domain via the $\mathbb{Z}_2$ transformations, before applying the ortho-normalisation condition. 
The 4D effective Yukawa coupling constants are then changed by an additional factor $1/2$, as is clear from the dimensional wave function normalisation
forms~\eqref{Mix_KK_Normalization_S1/Z2_R__Func_Yuk}-\eqref{eq:orthonorm-on-interv},  
which confirms the result~\eqref{Couplings_S1/Z2_R_Yuk_5D_BBT}. On the other side, we see as well that the fermion masses so obtained (unchanged by the spatial 
domain extension) involve only a new normalised parameter $X/2$, with respect to Ref.~\cite{Angelescu:2019viv}, 
which confirms the found spectrum~\eqref{MS_S1/Z2_R__Func_Yuk_5D_BBT_2}.

Beyond these action correspondences, there are other elegant similarities. For example, as illustrated by Fig.~\ref{fig:Pyramidal}, both the interval and orbifold scenarios lead to
the same vanishing probability current conditions at the two branes (and hence to identical EBC); those current conditions come, respectively, directly from the interval boundary 
criteria and indirectly from $\mathbb{Z}_2$ symmetry considerations. Besides, Table~\ref{ST_S1/Z2_R__Func} 
shows that the same treatments of the two branes, at the fixed points or interval boundaries, must be adopted in identical situations and that the same BC are generated.

Finally, let us propose an intuitive description for understanding the orbifold versus interval model duality. The obtained wave functions for the bulk fermions on the interval are of the 
kind $\cos(M_n \, y) \propto (e^{iM_n \, y}+e^{-iM_n \, y})$, coming in factor (via the KK decomposition) of the energy coefficients $e^{\pm iE\, t}$ in the 4D Dirac fields, which gives rise to wave planes propagating in both 
$y$-directions of the interval with momenta $\pm p_n=\pm M_n$ -- as for oscillations left-moving and right-moving along opposite directions in the world-sheet parameter space of strings. The associated particle,   
going in the direction $L\to 0$ and then coming back along $0 \to L$, reproduces the propagation along $\mathcal{S}^1$, following consecutively the two fundamental domains 
$-\pi R\to 0^{-}$ and $0^{+}\to \pi R$ of the orbifold (effectively equivalent orientations of the circle in the bulk so a unique propagation direction chosen along it): exactly the same $\mathcal{L}\left[\Phi(x^\mu, y)\right]$
Lagrangian evolution is felt by this particle during those dual travelings along the extra $y$-dimension, in the two different models, as is clear from the Lagrangian $\mathbb{Z}_2$ symmetry 
depicted in the drawing~\ref{S1_Z2_R}.

\section{Conclusions}
\label{Conclusion}

In the study of the $\mathcal{S}^1/\mathbb{Z}_2$ orbifold, 
the proper action definition through improper integrals has allowed to obtain consistent bulk profile solutions with possible discontinuities at the fixed points.  
In particular the point-like interaction of Yukawa creates a profile jump. 

These solutions have been obtained without brane-Higgs regularisation, by relying on the necessary EBC, coming from vanishing fermion probability currents, 
or alternatively on the introduction of BBT in the action. The associated calculations have been confirmed by the matching, between the 4D and 5D approaches, of the analytical results 
for the fermion mass spectrum and 4D effective Yukawa coupling constants.

The orbifold version, with $\mathbb{Z}_2$ transformations of the fields extended to the fixed points, was shown to be able to generate the chiral nature of the theory 
and even to select the expected SM chirality configuration for the 4D states. 

The duality between the interval and orbifold scenarios has been deeply described. It has also constituted the opportunity to point out an alternative method for
calculating the tower of excitation masses and 4D Yukawa couplings.

We are now working on the introduction of distributions in this context~\cite{WIP}.

\vspace{1cm}

\noindent {\bf Acknowledgments}
\\ \\
\noindent R.L. is supported by the agreement signed between the {\it China Scholarship Council (CSC)} and the {\it Universit\'e Paris-Saclay}.

\newpage

\appendix

\section*{Appendix}

\section{Notations \& conventions}
\label{notations_and_conventions}

Throughout the paper, we use the conventions of Ref.~\cite{Schwartz:2013pla}.
The 5D Minkowski metric is,
\begin{equation}
\eta_{MN} = \text{diag}(+1, -1, -1, -1, -1) \, ,
\nonumber%\label{metric_1}
\end{equation}
where $M,N=0,1,...,4$.

The 4D Dirac matrices are taken in the Weyl representation,
\begin{equation}
\gamma^\mu =
\begin{pmatrix}
0 & \sigma^\mu \\
\bar{\sigma}^\mu & 0
\end{pmatrix}
\phantom{000} \text{with} \phantom{000}
\left\{
\begin{array}{r c l}
\sigma^\mu &=& \left( \mathbb{I}, \sigma^i \right) \, , \\
\bar{\sigma}^\mu &=& \left( \mathbb{I}, -\sigma^i \right) \, ,
\end{array}
\right.
\label{gamma_1}
\end{equation}
where $\mu=0,1,2,3$ and $\sigma^i$ ($i = 1, 2, 3$) are the three Pauli matrices:
\begin{equation}
\sigma^1 =
\begin{pmatrix}
0 & 1 \\
1 & 0
\end{pmatrix} \, ,
\phantom{000}
\sigma^2 =
\begin{pmatrix}
0 & -i \\
i & 0
\end{pmatrix} \, ,
\phantom{000}
\sigma^3 =
\begin{pmatrix}
1 & 0 \\
0 & -1
\end{pmatrix} \, .
\nonumber
\end{equation}

One has also the 4D chirality operator,
\begin{equation}
\gamma^5 = i \prod_{\mu = 0}^3 \gamma^\mu =
\begin{pmatrix}
- \mathbb{I} & 0 \\
0 & \mathbb{I}
\end{pmatrix} \, .
\label{gamma_2}
\end{equation}
In our conventions, the 5D Dirac matrices $\Gamma^M$ ($M=0,1,...,4$) obey $\left\{\Gamma^A, \Gamma^B \right\} = 2 \, \eta^{AB}$ ($A,B = 0,1,...,4$) and read as,
\begin{equation}
\Gamma^M = \left( \gamma^\mu, -i \gamma^5 \right) \, .
\label{gamma_3}
\end{equation}

%((\section{Types of $\mathbb{Z}_2$ transformations}
%\label{Z2_transfo}
%\input{Z2_transfo}))

%((\section{Distribution theory on $S^1$}
%\label{distribution_theory}
%\input{maths_distributions.tex}))

\section{From spinor components to compact notations}
\label{C2C}

\subsection{Spinor components and their variations}
\label{app:Spin.1}

The generic spinor field $F$ ($F=Q,D$) introduced via Eq.~\eqref{1_5DDiracSp} can be written in terms of its four explicit components
$F_\alpha$ [$\alpha=1,2,3,4$]:
\begin{equation}
F =
%\begin{pmatrix}
%\mathcal{F}_L \\
%\mathcal{F}_R
%\end{pmatrix}
%=
%\begin{pmatrix}
%\begin{pmatrix}
%\mathcal{F}_{L1} \\
%\mathcal{F}_{L2}
%\end{pmatrix} \\
%\begin{pmatrix}
%\mathcal{F}_{R1} \\
%\mathcal{F}_{R2}
%\end{pmatrix}
%\end{pmatrix}
%=
\begin{pmatrix}
F_1 \\
F_2 \\
F_3 \\
F_4
\end{pmatrix}
\, ,
\label{Sp_C2C}
\end{equation}
and similarly, $\bar F$ can be expressed in terms of its own four components $\bar F_\alpha$:
\begin{equation}
\bar{F}
=
\begin{pmatrix}
\bar{F}_1, &
\bar{F}_2, &
\bar{F}_3, &
\bar{F}_4
\end{pmatrix}
\hat = \, 
F^\dagger \gamma^0
%=
%\begin{pmatrix}
%\mathcal{F}_{R}^\dagger &
%\mathcal{F}_{L}^\dagger
%\end{pmatrix}
=
\begin{pmatrix}
F^*_3, &
F^*_4, &
F^*_1, &
F^*_2
\end{pmatrix}
\, .
\label{Sp_bar_C2C}
\end{equation}
These 8 components constitute the fundamental variables of the Lagrangian~\eqref{S_Psi_S1/Z2_R__Func}. 
Hence, the variation of the associated Action, $S_{bulk}$ [see Eq.~\eqref{eq:Sstart}], involves the following 8 elementary variations, that
we can group into new 4-component (transposed) vectorial objects defined as:
\begin{equation}
\delta F \, \hat =
\begin{pmatrix}
\delta F_1 \\
\delta F_2 \\
\delta F_3 \\
\delta F_4
\end{pmatrix}
\, , 
\ \ \ \delta\bar{F}
\, \hat =
\begin{pmatrix}
\delta\bar{F}_1, &
\delta\bar{F}_2, &
\delta\bar{F}_3, &
\delta\bar{F}_4
\end{pmatrix}
\, ,
\label{Sp_group}
\end{equation}
introducing the 8 components $\delta F_\alpha$ and $\delta \bar F_\alpha$. We then define,
\begin{equation}
\delta F  =
\begin{pmatrix}
\delta F_L \\
\delta F_R
\end{pmatrix}
, \
\delta\bar{F} \, \hat =
\begin{pmatrix}
\delta{F}^\dagger_R, &
\delta{F}^\dagger_L
\end{pmatrix}
 , \  \rm{with \ for \ instance,} \ 
\delta{F}^\dagger_R \,  \hat =
\begin{pmatrix}
\delta\bar{F}_1, &
\delta\bar{F}_2
\end{pmatrix}
\, ,
\label{Sp_group-2comp}
\end{equation}
inspired by the following generic relations, based on Eq.~\eqref{1_5DDiracSp},
\begin{equation}
F  =
\begin{pmatrix}
F_L \\
F_R
\end{pmatrix}
 , \, 
\bar{F} \, \hat = \, F^\dagger \gamma^0 = (F^\dagger_R,\, F^\dagger_L) \, .
\label{Sp_group-2comp-bis}
\end{equation}

\subsection{A typical compact form calculation}
\label{app:Spin.2}

Using the Lagrangian $\mathcal{L}_{kin}$ of Eq.~\eqref{S_Psi_S1/Z2_R__Func}, let us work out explicitly the following quantity
entering Eq.~\eqref{S_Psi_S1/Z2_R__Func_HVP} in a compact form (no explicit spinor index of type $\alpha$),
\begin{eqnarray}
 \delta {\bar{F}}  \,  \dfrac{\partial \mathcal{L}_{kin}}{\partial \bar{F}}
& \, \hat = & \sum_{\alpha=1}^4 \delta \bar{F}_\alpha \dfrac{\partial \mathcal{L}_{kin}}{\partial \bar{F}_\alpha}
 =  \sum_{\alpha=1}^4 \delta \bar{F}_\alpha \dfrac{\partial }{\partial \bar{F}_\alpha} 
\left ( \frac{i}{2} \, \bar F \Gamma^{M} \partial_{M} F \right ) 
\nonumber \\ & = &  \sum_{\alpha=1}^4 \delta \bar{F}_\alpha \dfrac{\partial }{\partial \bar{F}_\alpha} 
\left ( \frac{i}{2} \, \sum_{\beta=1}^4 \bar F_\beta [\Gamma^{M} \partial_{M} F]_\beta \right ) 
= \sum_{\alpha=1}^4 \delta \bar{F}_\alpha  \frac{i}{2} \, [\Gamma^{M} \partial_{M} F]_\alpha 
\nonumber \\ & = &  \frac{i}{2} \, \delta \bar F \Gamma^{M} \partial_{M} F  \, ,
\label{app:CompCalc}
\end{eqnarray}
where the spinor components of Eq.~\eqref{Sp_C2C} and \eqref{Sp_bar_C2C} have appeared, as well as the
variations of Eq.~\eqref{Sp_group}.

\subsection{$\mathbb{Z}_2$ transformations of field variations}
\label{app:Spin.3}

Finally, we can derive the $\mathbb{Z}_2$ transformation for the compact form $\delta \bar{F}$ of Eq.~\eqref{Sp_group}.
Accordingly to the $\mathbb{Z}_2$ transformations~\eqref{Parity_S1/Z2_R_1}-\eqref{Parity_S1/Z2_R_4}, we have,
\begin{align}
\bar{F}|_{-y} = F^\dagger|_{-y} \gamma^0 =
\left( \pm   \gamma^5 F\right)^\dagger|_{y} \gamma^0 = 
\pm   F^\dagger|_{y} \gamma^5 \gamma^0 = \mp   F^\dagger|_{y} \gamma^0 \gamma^5 = \mp   \bar{F}|_{y} \gamma^5 \, ,
\nonumber%\label{Parity_Sp_bar_S1/Z2_R}
\end{align} 
due to the anti-commutator relation $\left\{\gamma^5, \gamma^\mu \right\}=0$. Then one must rewrite this relation by making the
spinor components of Eq.~\eqref{Sp_bar_C2C} appear explicitly:
\begin{align}
\bar{F}_\alpha |_{-y} = \mp   \sum_{\beta=1}^4  \bar{F}_\beta |_{y} \gamma^5_{\beta \alpha} \, , \nonumber
\end{align} 
in order to deduce the relation on the variations of these components:
\begin{align}
\delta \bar{F}_\alpha |_{-y} = \mp   \sum_{\beta=1}^4  \delta \bar{F}_\beta |_{y} \gamma^5_{\beta \alpha} \, . \nonumber
\end{align} 
Thanks to Eq.~\eqref{Sp_group}, this equation can be contracted back to the compact notation as,
\begin{align}
\delta \bar{F}|_{-y} = \mp   \delta \bar{F}|_{y} \gamma^5 \, .
\label{Parity_Delta_Sp_bar_S1/Z2_R}
\end{align}

\newpage

%\section{Glossary}
%\label{Glossary}
%\input{glossary.tex}

%\newpage

\bibliographystyle{JHEP}

%\bibliography{bib,bibli,bib_metric_graphs,bib_DM,bib_neutrinos,bib_ADD,bib_BSM,bib_Classicalization,bib_Divers,bib_ED,bib_Exp,bib_GHU,bib_gravity,bib_HiggsFits,bib_LQG,bib_Maths,bib_QFT,bib_RS,bib_SM,bib_Strings,bib_GM}
%\bibliography{S1_Z2.GM}
\bibliography{S1_Z2_R__Func}

\end{document}